\documentclass[11pt]{article}

\usepackage[T1]{fontenc}
\usepackage{type1cm}

\usepackage{authblk}

\usepackage{amsmath}
\usepackage{amssymb}
\usepackage{amsthm}
\usepackage{dsfont}

\usepackage{booktabs}
\usepackage{graphicx}
\usepackage{subcaption}
\usepackage{multirow}
\usepackage{adjustbox}
\usepackage[section]{placeins}
\usepackage{flafter}

\usepackage[numbers,sort&compress]{natbib}
\usepackage{hyperref}
\usepackage{cleveref}
\usepackage{xurl}
\usepackage{siunitx}
\usepackage{fullpage}
\usepackage{dsfont}

\usepackage{xcolor}
\hypersetup{
colorlinks=true,
linkcolor={red!50!black},
filecolor={green!50!black},
citecolor={green!50!black},
urlcolor={blue!80!black},
}

\title{Conspiracy and Environment Communities on Reddit\\Sort Along Different Demographic Axes}

\author[1,2]{%
Miguel \'{A}. S\'anchez-Cort\'es%
\thanks{Corresponding author: \href{mailto:sanchezm@ceu.edu}{sanchezm@ceu.edu}. %
}%
}

\author[3]{Corrado Monti}

\author[4]{Gianmarco~De~Francisci~Morales}

\affil[1]{%
Department of Network and Data Science,
Central European University,
Quellenstra{\ss}e 51,
1100 Vienna, Austria
}

\affil[2]{%
Department of Computer Science,
Sapienza University of Rome,
Viale Regina Elena 295,
00161 Rome, Italy
}

\affil[3]{%
Department of Humanities,
University of Turin,
via S. Ottavio 20,
10124 Turin, Italy
}

\affil[4]{%
CENTAI,
Corso Inghilterra 3,
10138 Turin, Italy
}

\date{}

\begin{document}

\maketitle

\begin{abstract}
The impact of online platforms in political mobilization is widely documented, yet it remains unclear how it affects different demographic groups.
To shed light on this phenomenon, we examine two contrasting cases: engagement with environmental causes on one side, and with conspiracy theories on the other.
Both are well represented on Reddit, allowing us to study the different origins of these groups using a network-based method.
By inferring age, gender, affluence, and partisanship of Reddit users, we construct stratified transition networks to compare the entry pathways of different demographic groups toward these communities.
We find that demographic differentiation operates differently in these two domains.
Entry into conspiracy communities is primarily structured by gender and partisanship: masculine users disproportionately arrive through overtly political and alt-right spaces and share right-wing media sources, whereas feminine users more often pass through esoteric and spiritual subreddits before converging on r/conspiracy.
In contrast, environmental communities are differentiated mainly by age and affluence.
Affluent users tend to approach environmental subreddits via discussions of individual energy management, technology, and finance, while less affluent users arrive through protest-oriented and climate movement spaces.
By comparing these two issue domains, the study shows that sociodemographic sorting does not operate uniformly across polarized topics.
Instead, distinct demographic groups build their own paths, shaping both the routes through which they engage with political content and the informational ecosystems that sustain their engagement.
Our approach, based on the attention-flow graph, refines our understanding of how different groups become embedded in online political communities.
\end{abstract}

\noindent\textbf{Keywords:}
social media, network science, conspiracy, sociodemographics

\section{Introduction}\label{sec:intro}

Digital platforms have become a central infrastructure for political mobilization and opinion formation, where individuals exchange information, discuss public issues, and take early, observable steps toward collective engagement~\citep{boulianneTwentyYearsDigital2020}. 
Online political communities are often studied as spaces where polarization and segregation are already visible~\citep{pariserFilterBubbleWhat2011, cinelliEchoChamberEffect2021a,montiEvidenceDemographicRather2023}.
This view, however, leaves open a prior question: how do individuals enter these spaces in the first place?

This question takes on particular significance once entry is understood as a demographically structured process: political conflict is increasingly organized around the alignment of demographic identities, social positions, and partisan affiliations, a process sometimes described as the ``Big Sort''~\citep{bishopBigSortWhy2008, masonUncivilAgreement2018}.
This structure extends to online environments: in Reddit news discussions, users are segregated more by age and affluence than by ideology~\citep{montiEvidenceDemographicRather2023}.
Yet this evidence is largely static, because it shows that demographic groups occupy different online spaces~\citep{wallerQuantifyingSocialOrganization2021} but not whether they traverse different communities and information environments before entering politically salient ones.
If demographic sorting shapes participation in political communities, it should also be visible upstream: in the pathways through which different groups enter these communities, and in the information environments they inhabit along the way.

We examine this question through two theoretically contrasting cases: conspiracy and environmental communities on Reddit.
Both communities have moved from online discussion to substantial political consequence: QAnon narratives have become entangled with electoral politics~\citep{pew5FactsQAnon2020, engelCharacterizingRedditParticipation2022, sharmaCharacterizingOnlineEngagement2022}, while climate concern has shaped voting behavior among younger generations~\citep{wahlstromSurveysParticipantsFridays2020, neasYoungPeoplesClimate2022, YouthConcernClimate2022, UnderstandingProclimateVoters}.
Reddit's role in incubating each movement is independently documented, from early QAnon-sympathetic subreddits~\citep{engelCharacterizingRedditParticipation2022} to the founding of the Earth Strike movement following a Reddit post~\citep{lentiCausalModelingClimate2025}.

Beyond their individual salience, the analytical value of this comparison lies in the contrast between them.
Drawing on social movement theory, we treat participation in online communities as an early stage of collective engagement and distinguish between two ideal-typical logics: issue-oriented engagement, organized around shared policy goals, and identity-expressive engagement, organized around shared belonging~\citep{klandermansDemandSupplyParticipation2004, pollettaCollectiveIdentitySocial2001, vanstekelenburgSocialPsychologyProtest2013}.
Environmental communities align more closely with issue-oriented engagement, as they focus on a public issue, climate change, and competing forms of environmental action grounded in institutional science and expert assessment~\citep{fage-butlerPublicTrustMistrust2022}.
Conspiracy communities align more closely with identity-expressive engagement, as they center on distrust of mainstream institutions, alternative epistemic authority, and diffuse political suspicion~\citep{delvicarioSpreadingMisinformationOnline2016,douglasUnderstandingConspiracyTheories2019, uscinskiPsychologicalPoliticalCorrelates2022}.
The two cases also differ in ways that are central to our demographic question: environmental activism is often associated with younger cohorts, students, and highly educated publics, while conspiracy beliefs have been linked to political identity, institutional distrust, social marginalization, and perceived lack of control~\citep{wahlstromSurveysParticipantsFridays2020, neasYoungPeoplesClimate2022, douglasUnderstandingConspiracyTheories2019, uscinskiPsychologicalPoliticalCorrelates2022}.
This distinction is not absolute---climate activism also produces collective identities, and conspiracy communities can mobilize around concrete issues---but it generates a useful analytical contrast between communities that differ in their relationship to public problems and institutional authority.
These logics suggest that demographic sorting may appear at different stages in the pathway: upstream for identity-expressive communities, where prior affiliations organize access, and downstream for issue-oriented communities, where different groups engage with the same public problem through different repertoires.

We operationalize entry as a user's first post in conspiracy and environmental communities, and trace the demographic pathways that precede it using the attention-flow graph framework~\citep{rolloCommunitiesGatewaysBridges2022}, which models how users shift their attention from one subreddit to another over time.
This framework captures a dynamic aspect of online community participation that surveys and static analyses do not typically capture: the information environments users traverse before entering a political community.
Specifically, the framework allows identifying \emph{gateways} (subreddits that act as entry points into a community) and \emph{bridges} (subreddits that facilitate transitions between communities).
We extend the attention-flow graph with sociodemographic stratification by incorporating the axes of age, gender, affluence, and partisanship defined by \citet{wallerQuantifyingSocialOrganization2021}.

We find that demographic sorting does not operate uniformly across the two cases, and that the stage at which demographic groups diverge differs in ways congruent with the theoretical contrast.
Consistent with an identity-expressive logic, entry into conspiracy communities is structured primarily upstream, especially by gender and partisanship.
In fact, masculine and right-leaning users arrive through overtly political and alt-right spaces circulating right-wing media sources, while feminine users more often pass through esoteric subreddits before converging on conspiracy communities.
In contrast, consistent with an issue-oriented logic, entry into environmental communities is structured primarily downstream, especially by age and affluence.
Affluent and older users approach environmental subreddits through discussions of energy, technology, and finance, while less affluent and younger users arrive through protest-oriented and climate-movement spaces.
These results show that demographic sorting is not only a matter of who ends up in political communities, but of how different groups get there, and that they do so through distinct information environments.
More broadly, these results suggest that the information environments different groups inhabit before entering political communities are themselves demographically segregated, and that the form this segregation takes depends on the type of political engagement at stake.

\section{Results}
\begin{figure*}[!t]%
\centering
\includegraphics[width=0.7\linewidth]{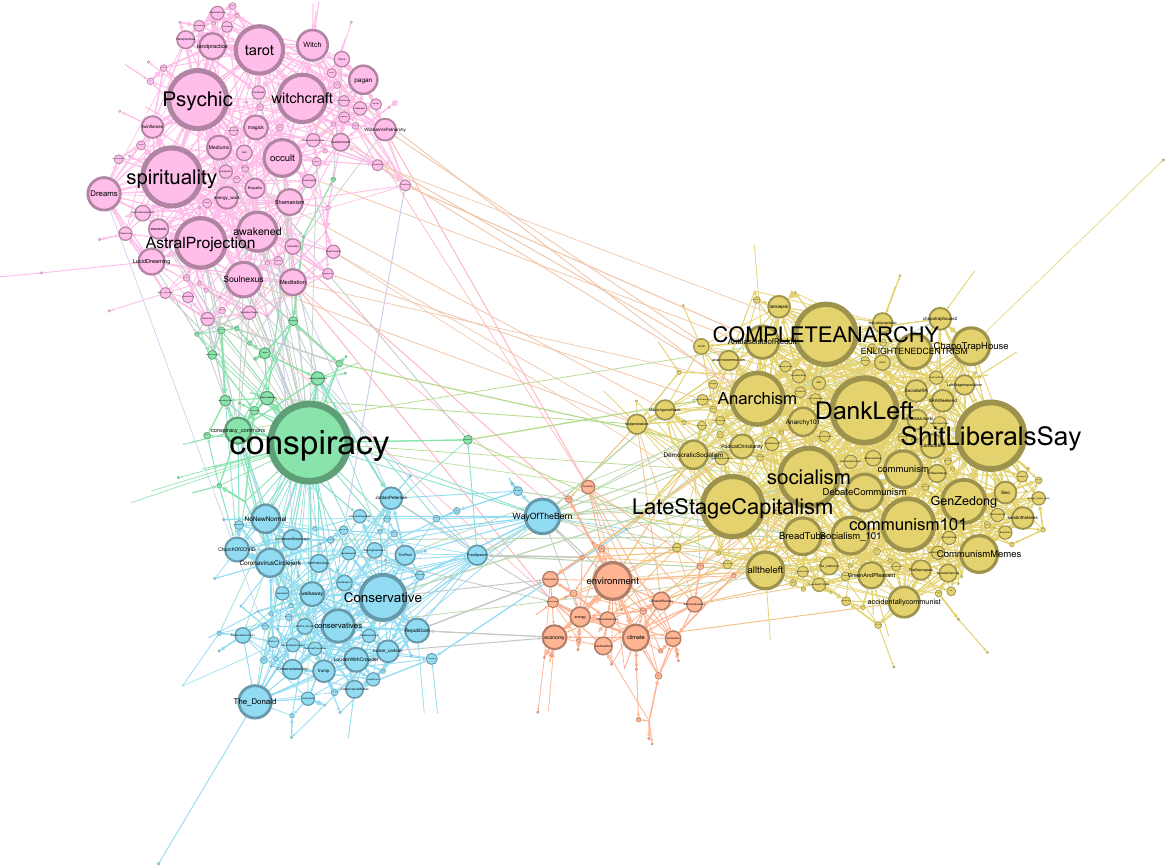}
\caption{\textbf{Political and conspiracy communities in the Attention Flow Graph.}
Subgraph of the Attention Flow Graph highlighting selected political and conspiracy-related Reddit communities.
Node colours indicate community membership: Conspiracy (green), Esoterism (pink), Alt-Right (blue), Radical Left (yellow), and Environment (orange).
Node size is proportional to subreddit in-degree, and edge width is proportional to edge weight.}
\label{fig:communities}
\end{figure*}
\begin{figure*}
\centering
\includegraphics[width=\linewidth]{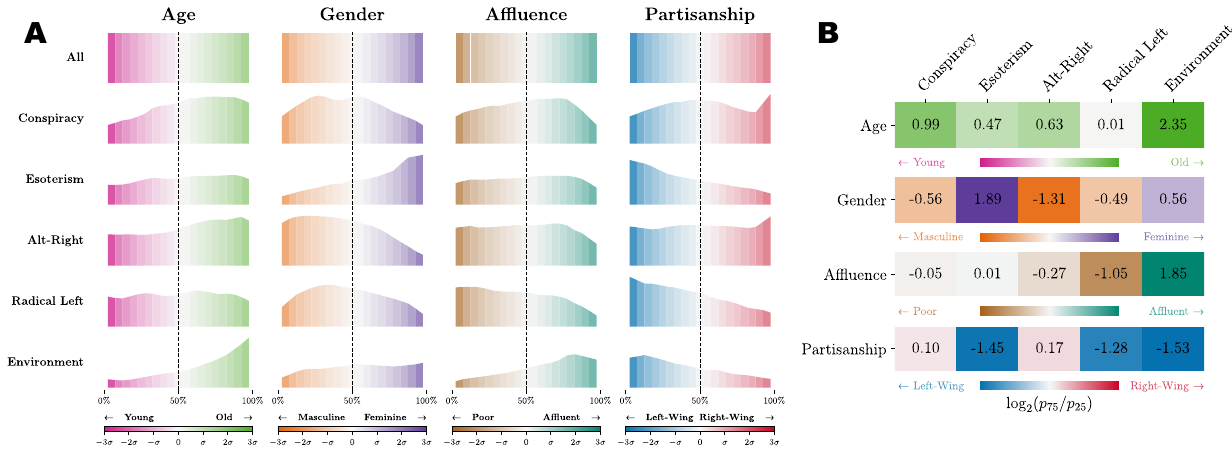}
\caption{\textbf{Sociodemographic composition of community entrants and group overrepresentation.}
\textbf{A.} Distributions of sociodemographic scores of users entering each community.
\textbf{B.} Log-likelihood ratios comparing the 75th and 25th percentiles of the distributions, highlighting overrepresented groups.}
\label{fig:combined-quantile-llr}
\end{figure*}

We first contextualize environmental and conspiracy communities within the broader Reddit political ecosystem.
Then, we analyze their demographic makeup to understand the specific sociodemographic biases present.
Finally, we trace the distinct pathways demographic groups take to join these communities, exploring their journeys across macroscopic, mesoscopic, and microscopic scales.

\subsection{Attention flow graph}

\Cref{fig:communities} shows the attention flow graph (AFG) induced by user migration across Reddit communities.
In the AFG, nodes are subreddits and weighted directed edges capture the chronological migration of active users from one community to another~\citep{rolloCommunitiesGatewaysBridges2022} (see \nameref{sec:methods}).
The graph highlights how our target communities are embedded in the broader political ecosystem on Reddit, considering the same political context identified by \citet{rolloCommunitiesGatewaysBridges2022}.
We see that the Conspiracy community is positioned between the  Esoteric and Alt-Right clusters, suggesting hybrid entry pathways.
Environment, on the other hand, sits between the more explicitly political subreddits, close to the Radical Left and Alt-Right communities.

\subsection{Demographic composition}

We now turn to the specific demographic composition of the communities under study.
\Cref{fig:combined-quantile-llr} presents two different views of the estimated composition of each community.
\Cref{fig:combined-quantile-llr}{A} shows the distribution of sociodemographic scores for users entering each community, and, in particular, its departure from the uniform reference distribution of the Reddit population (top row).
Thus, these score distributions need to be interpreted by keeping the Reddit demographics in mind.
For instance, by comparing the age scores with self-declared age from users, we observe that the median age score corresponds to a self-declared age of 23 (see \Cref{fig:age-comparison}).
The distributions indicate that entrants to Conspiracy are relatively older, more masculine, more affluent, and more right-leaning than the Reddit baseline.
Entrants to Environment are also relatively older and more affluent, but are comparatively more left-leaning.

\Cref{fig:combined-quantile-llr}{B} summarizes these patterns using the log-likelihood-ratio between the 75th and 25th percentiles.
This statistic quantifies which side of each sociodemographic axis is over-represented among entrants.
The strongest effects appear for Environment (older, affluent, left-leaning), Esoterism (feminine, left-leaning), and Alt-Right (masculine).
By contrast, Conspiracy shows comparatively weaker aggregate skew, with the clearest signal on age.

\subsection{Demographic Pathways}

\begin{figure*}[!t]
\centering
\includegraphics[width=\linewidth]{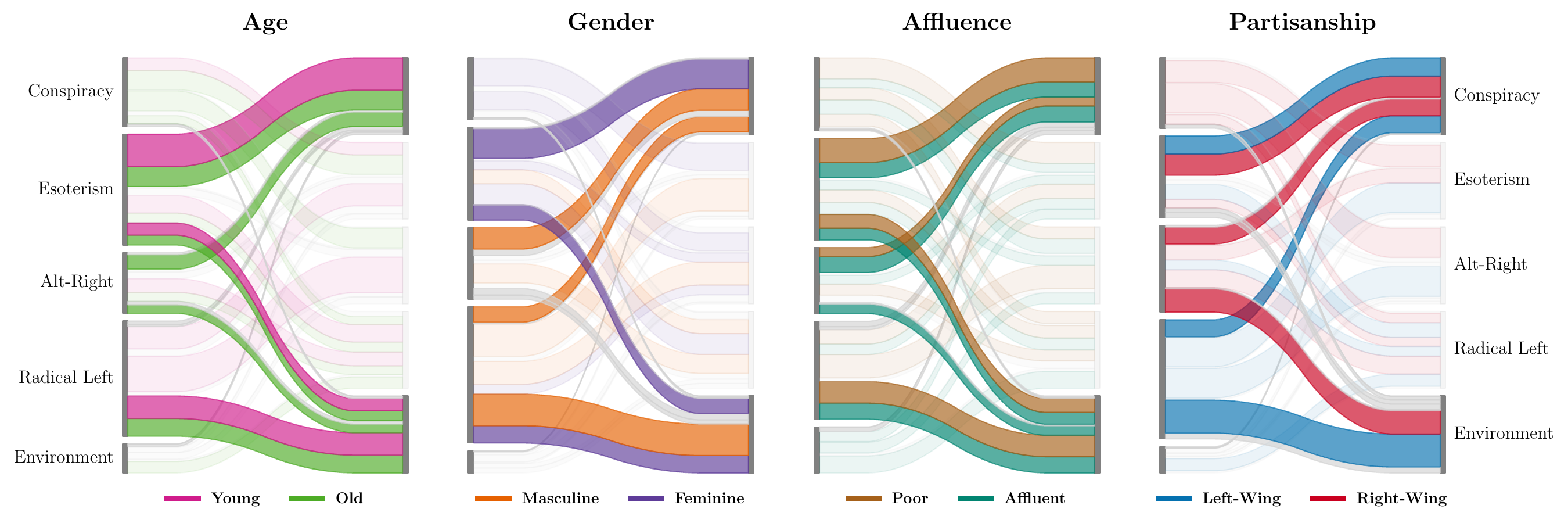}
\caption{\textbf{Cross-community transition probabilities by sociodemographic group.}
Probability that a random walk ending in a given right-side community originates from each left-side community, stratified by demographic group.
Flows terminating in the two focal communities, \emph{Environment} and \emph{Conspiracy}, are highlighted.}
\label{fig:sankey}
\end{figure*}
\begin{figure*}
\centering
\includegraphics[width=0.8\linewidth]{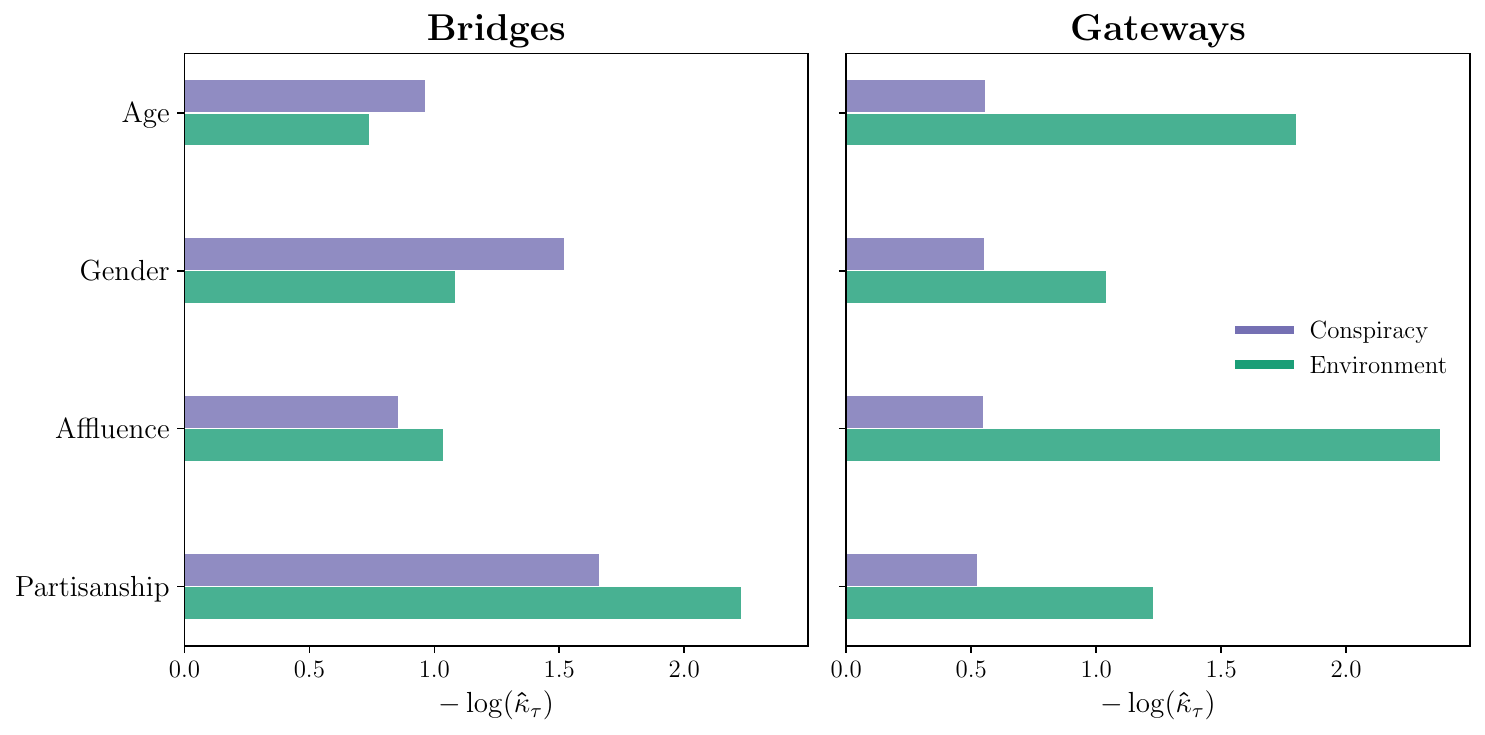}
\caption{\textbf{Differences in bridge and gateway rankings for conspiracy and environment communities.}
Negative logarithm of the weighted Kendall--Tau ($\hat{\kappa}_{\tau}$) values between the bridge and gateway rankings obtained from the positive and negative variants of the S-AFG for each social dimension.
Higher values indicate greater ranking differences between sociodemographic groups for Conspiracy and Environment.
}
\label{fig:kendall-tau-barplot}
\end{figure*}

Let us now focus exclusively on our two communities of interest, Conspiracy and Environment.
We use these communities to refine our research question further: do different demographics join these communities through different paths?

We answer this question from several angles by gradually zooming in on user behavior across three scales of resolution:
macroscopic flow between broad communities, mesoscopic flow through specific gateway subreddits, and microscopic flow of external content (URLs).
To answer these questions from such angles, we use the Stratified Attention Flow  (S-AFG).
This graph is a network representation of how users of different sociodemographic groups flow across communities, and it is a direct extension of the original AFG (see \nameref{sec:methods}).

\subsubsection{Community-level random walks}
To better understand the general flow of users, we use a random walk with restart to compute the probabilities of a user arriving at Conspiracy/Environment starting from the other communities we consider.
\Cref{fig:sankey} shows the results for this analysis.
Masculine users arrive at Conspiracy differently than feminine users: the former arrive from political communities---especially Alt-Right---while the latter arrive from esoteric communities.
For Environment, instead, this gender-based demographic split is not apparent: the largest influx of users is from Radical Left, and it affects both masculine and feminine users.
Regarding affluence, we do not observe significant self-sorting at this level of granularity: we shall show how a subreddit-level analysis points to different entry points on this demographic axis.
The partisanship axis shows a clear split of origins, aligned with the political orientation of the communities, as one might expect.
Interestingly, while the Esoterism community is predominantly left-wing, there is a considerable flow of right-wing users who join Conspiracy from it.
This result highlights the additional fine-grained information that such a dynamic analysis allows to gather.
As already noted, there is no politically-aligned self-sorting at the endpoints, and both orientations are well-represented in both Conspiracy and Environment.

\subsubsection{Bridges and Gateways}
We further quantify these differences in \Cref{fig:kendall-tau-barplot} by comparing separate versions of the S-AFG for each social dimension (e.g., isolating high vs. low affluence, or masculine vs. feminine users, see \nameref{sec:methods}).
We find the largest difference in the bridges (outside the community) for Conspiracy, and in the gateways (inside the community) for Environment.
Additionally, they are segregated by different axes: Conspiracy is differentiated by gender and partisanship, while Environment is by age and affluence.

These results suggest that user differentiation occurs at different stages of the pathway.
For Conspiracy, heterogeneity is mainly \emph{upstream}, before entry (bridges), especially by gender and partisanship.
For Environment, it is mainly \emph{downstream}, after entry (gateways), mainly by affluence and age.

\begin{table*}[!t]
\caption{\textbf{Top bridge and gateway communities for conspiracy and environmental content.}
Top-10 bridge communities ranked by personalized PageRank (PPR) for Conspiracy on the gender axis (left), and top-10 gateway communities for Environment on the affluence axis (right).
We exclude subreddits with fewer than 5 average unique users per month to avoid noise.\label{tab:top-bridges-gateways}}
\tabcolsep=0pt%
\begin{tabular*}{\textwidth}{@{\extracolsep{\fill}}ll ll@{\extracolsep{\fill}}}
\toprule%
\multicolumn{2}{@{}c@{}}{Top-10 Bridges (Conspiracy)} & \multicolumn{2}{c@{}}{Top-10 Gateways (Environment)} \\[2pt]
\cline{1-2}\cline{3-4}\\[-8pt]%
Masculine & Feminine & Poor & Affluent \\[-2pt]
\midrule
r/CoronavirusCirclejerk & r/Retconned & r/climate & r/environment \\
r/Paranormal & r/Wicca & r/climatechange & r/energy \\
r/ChapoTrapHouse & r/AstralProjection & r/ClimateOffensive & r/ZeroWaste \\
r/Conservative & r/awakened & r/ExtinctionRebellion & r/business \\
r/JordanPeterson & r/Psychic & r/economy & r/Economics \\
r/ShitPoliticsSays & r/Soulnexus & r/ClimateActionPlan & r/economy \\
r/chapotraphouse2 & r/tarot & r/AnimalRights & r/sustainability \\
r/SocialistRA & r/LucidDreaming & r/environment & r/climate \\
r/communism & r/CoronavirusCirclejerk & r/sustainability & r/RenewableEnergy \\
r/leftistvexillology & r/SpiritualAwakening & r/RenewableEnergy & r/climatechange \\
\bottomrule
\end{tabular*}
\end{table*}

\begin{figure*}
\centering
\makebox[\textwidth]{%
  \includegraphics[width=1\linewidth]{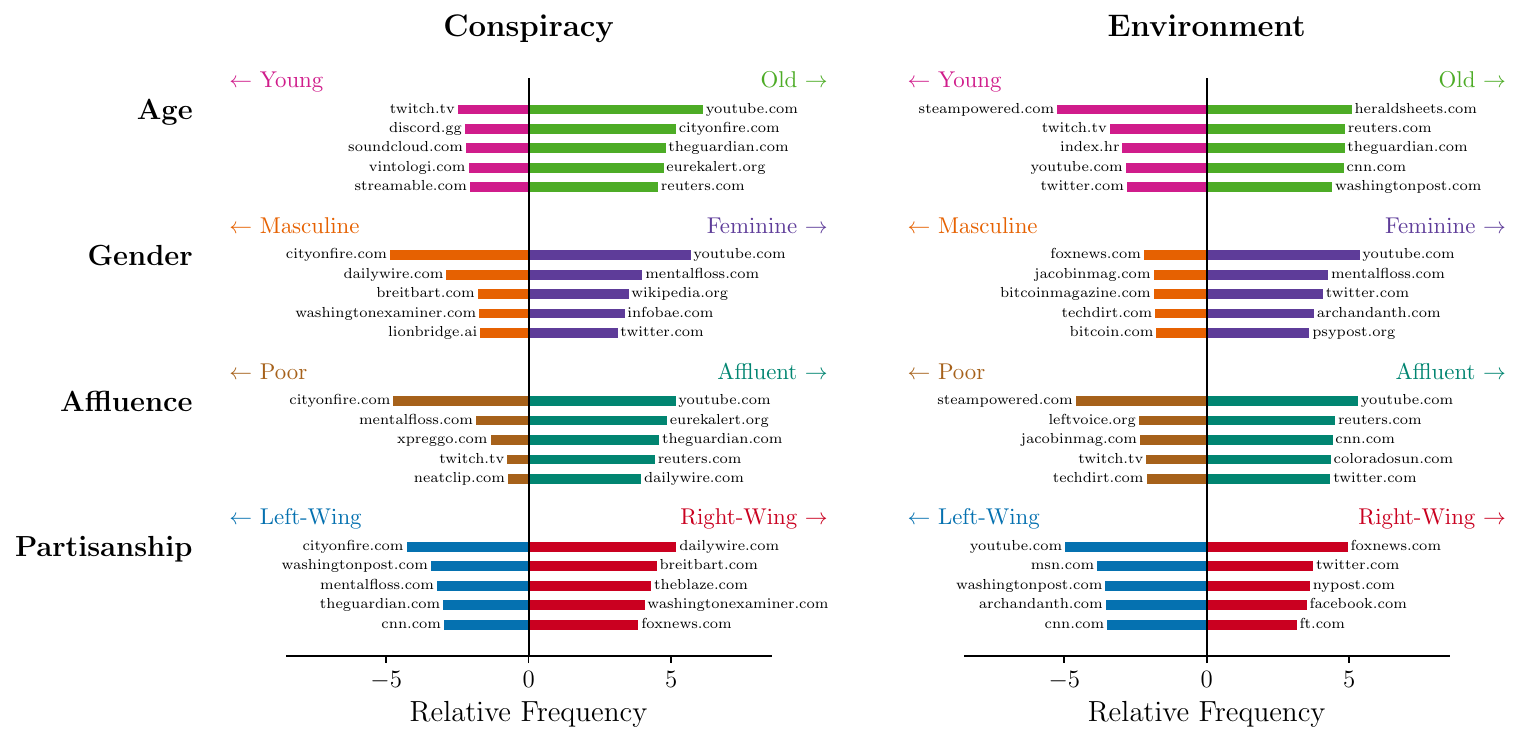}
}
\caption{\textbf{Most shared URL domains prior to entry into conspiratorial and environmental communities.}
Top five most shared URL domains by sociodemographic group before entering conspiratorial (left) and environmental (right) communities.
For each social dimension $d$, the ranking $r_i^d$ of domain $i$ is computed as a weighted sum of domain-sharing frequencies across users, weighted by their score in dimension $d$, followed by a logarithmic transformation to reduce the influence of outliers and skewed distributions.
Results are reported for users within the first and third quartiles of each social dimension to reduce noise.}
\label{fig:url_sharing_filtered}
\end{figure*}

To interpret these differences qualitatively, we examine the mesoscopic level: the specific communities that act as bridges and gateways in \Cref{tab:top-bridges-gateways}.
We focus on the top bridges of Conspiracy and the top gateways for Environment.
This choice is motivated by the previous result, which shows that the entry points of Conspiracy are differentiated outside the community, particularly by gender.
Since they are outside the Conspiracy community, they are defined as bridges, while inside the community most users converge to r/conspiracy.

The top bridges for Conspiracy by gender show that masculine users are more likely to pass through ideologically explicit political subreddits from both sides of the spectrum (e.g., r/ChapoTrapHouse and r/Conservative).
By contrast, feminine users disproportionately go through esoteric communities such as r/Retconned, r/Wicca, and r/AstralProjection, consistent with the flow patterns in \Cref{fig:sankey}.

Conversely, Environment shows significant differences in the subreddits used by different demographic groups \emph{within} the community, particularly along the affluence axis.
The gateway split by affluence suggests two distinct framings of climate engagement: affluent users are more connected to subreddits centered on energy management, sustainability practices, and economics, whereas less affluent users are more connected to protest and collective-movement-oriented communities.

\subsubsection{URL analysis.}
Let us further investigate the microscopic differences in behavior across these demographic splits by looking at the URLs shared by each group \emph{before} joining such communities.
Results are depicted in \Cref{fig:url_sharing_filtered}.
Here we see that masculine users join Conspiracy after sharing URLs from right-wing sources such as Breitbart and Daily Wire, and martial arts websites (\texttt{cityonfire.com}).
Feminine users instead join Conspiracy after sharing more generalist content, including YouTube videos, Wikipedia articles, and broad-interest news such as \texttt{mentalfloss.com}.

Regarding Environment, \Cref{fig:url_sharing_filtered} confirms that less affluent users come from protests and climate movements: sources such as Jacobin magazine and \texttt{leftvoice.org} appear among the most shared.
Interestingly, the video game platform Steam appears in the first place (possibly indicative of a younger population).
Instead, affluent users before joining Environment subreddits share more established sources: Twitter, YouTube videos, and news from CNN and Reuters.

\section{Discussion} \label{sec:discussion}

We have examined how demographic sorting affects the pathways of users into two key political communities on Reddit, the conspiratorial and the environmental ones.
As posited in the \nameref{sec:intro}, the choice of these two communities is not only due to their consequentiality on U.S. politics, but because they represent a theoretical distinction between two distinct types of political communities.
While the environmental community is based mostly on individuals interested in a specific issue~\citep{schlosbergEnvironmentalClimateJustice2014}, the conspiratorial community is more rooted into the personal identity of its users, since it reflects a broader worldview structured around distrust of mainstream experts, institutions, and information sources~\citep{suttonConspiracyTheoriesConspiracy2020}.
Our primary goal was to assess whether demographic sorting into political communities is visible upstream, in the pathways users traverse before entry, rather than only in the composition of the communities themselves.
We find that demographic differentiation is present in both domains but operates along different axes and, crucially, at different stages of the user journey---in a way consistent with the theoretical contrast between identity-expressive and issue-oriented engagement.

In Conspiracy communities, sorting occurs primarily \emph{before} entry, in the bridges users traverse, and is organized by gender and partisanship. In Environment communities, sorting emerges primarily \emph{after} entry, in the gateways users occupy within the community, and is organized by age and affluence. The stage at which demographic groups diverge thus appears to reflect the underlying logic of the community: identity-based communities differentiate users by their path to belonging, while issue-based communities by their orientation toward action.

The upstream sorting into conspiracy communities is consistent with their identity-expressive character~\citep{pollettaCollectiveIdentitySocial2001}: what these communities offer is a shared identity of institutional distrust, which users can reach through distinct cultural pathways.
Masculine users arrive through explicitly political subreddits spanning both ends of the partisan spectrum---from r/ChapoTrapHouse and r/communism to r/Conservative and r/JordanPeterson---while circulating right-leaning media such as Breitbart and the Daily Wire; feminine users instead traverse esoteric and spiritual communities (r/Wicca, r/AstralProjection, r/tarot) while sharing more generalist content.
What unites these otherwise disparate routes is not a single ideology but a shared rejection of mainstream epistemic authority, reached through a partisan mechanism in one case and a spiritual one in the other.
Moreover, once inside, this diversity of routes converges on a shared belonging~\citep{sterniskoDarkSideSocial2020}: the internal gateway structure is comparatively undifferentiated, with all groups funneling into r/conspiracy.

Environmental communities display a different pattern, consistent with an issue-oriented logic~\citep{klandermansDemandSupplyParticipation2004}: demographic groups converge on the same public problem but diverge in the action repertoires through which they engage with it.
The gateways that distinguish groups here are concrete practices---r/ZeroWaste, r/RenewableEnergy, r/ClimateOffensive, r/ExtinctionRebellion---rather than identities.
The affluence divide maps onto a broader individualist--collectivist contrast in worldview: affluent and older users approach the issue through personal-responsibility repertoires (green consumption, sustainable energy) that are materially more accessible to those with the purchasing power to adopt them and that afford a sense of individual efficacy over the problem~
\citep{angIndividualismAdoptionClean2020}.
For less affluent and younger users, for whom these personal pathways are constrained, a collectivist framing that locates solutions in structural political change and protest movements is plausibly more resonant~\citep{choBeNotBe2013}.

\smallskip
A few limitations should be considered in our interpretation.
First, the comparison between communities is theoretically motivated but necessarily imperfect.
These communities provide a useful empirical contrast between identity-expressive and issue-oriented engagement, but they are not pure types: environmental politics can generate collective identities, and conspiracy communities can mobilize around concrete issues.
Our cases should therefore be read as a pragmatic compromise between theoretical clarity and the messiness of real online political ecosystems.
Second, our sociodemographic measures are only behavioral proxies.
As a result, they are most informative for users whose Reddit behavior is typical of their demographic.
While these proxies have been validated (by \citet{wallerQuantifyingSocialOrganization2021}, and in the present work, see section \ref{si:sociodemographic-scores}), they inevitably highlight the most stereotypical users.
Consequently, our results may understate within-group heterogeneity.
The patterns we report should therefore be interpreted as differences between demographically coded behavioral profiles, rather than as representative estimates for all members of a given demographic group.
Third, the pathways we observe are limited to visible activity on Reddit.
Users will encounter political content through other subreddits outside our filtered graph, through comments rather than submissions, through private messages, or through other platforms entirely.
The trajectories reconstructed here should therefore be understood as Reddit-visible pathways into these communities, not as complete histories of exposure, persuasion, or mobilization.
These limitations do not undermine the observed pattern of differential sorting, but they caution against treating the measured pathways as exhaustive.

These findings carry implications for how we understand self-sorting and polarization online. Prior work shows that demographic groups occupy different political spaces~\citep{wallerQuantifyingSocialOrganization2021, montiEvidenceDemographicRather2023}; we show that they also \emph{travel} through demographically segregated information environments before arriving there. However, our analysis is observational: pathways describe how groups arrive, not why, and the mechanisms we propose remain interpretive. Future work could test whether the sorting pattern generalizes to other identity-based and issue-based communities and platforms, whether entry pathway predicts subsequent trajectories (persistence, radicalization, or offline mobilization), and whether the upstream segregation we document causally shapes the beliefs users hold once inside. Linking pre-entry information environments to downstream political behavior would clarify whether these demographically distinct roads lead to genuinely different destinations.

\section{Methods} \label{sec:methods}

This section is organized as follows.
We begin by describing the Reddit dataset we construct.
We then detail how we infer users' sociodemographic profiles from their posting behaviour.
Next, we introduce the Stratified Attention Flow Graph (S-AFG), our core methodological contribution.
Finally, we describe the suite of measures computed on the (S-)AFG: community detection, sociodemographic profiling of community entrants, gateway and bridge identification, and shared-URL analysis.

\subsection{Dataset} \label{subsec:data}

We construct a novel dataset of recent Reddit activity comprising all posts from 2019 to 2023, drawn from the Pushshift dataset \citep{baumgartnerPushshiftRedditDataset2020}.
We focus exclusively on submissions rather than comments, as posts provide a stronger proxy for a user's \emph{involvement} in a subreddit \citep{dattaExtractingInterCommunityConflicts2019}.

Given the substantial growth in Reddit activity over time---which makes full-scale analysis computationally expensive for recent years---we draw a random 10\% sample of users who posted during the 2019--2023 period.
Because our analysis operates at the user level and concerns demographic characteristics, we sample uniformly across users rather than across activity, avoiding bias toward highly active individuals.
The 10\% sampling rate is chosen to balance representativeness with computational tractability.

We apply three further filters to ensure data quality.
First, to ensure meaningful engagement, we retain only posts with a score (upvotes minus downvotes) greater than one.
Second, to restrict attention to active communities, we include only subreddits in which at least 50 distinct users made a post in at least one calendar month.
Third, to mitigate bot activity, we exclude users who posted in more than 50 distinct subreddits within a single month and users whose usernames contain the string ``bot''.
The resulting dataset contains \num{50360619} posts from \num{4536459} unique users across \num{89157} subreddits.

\subsection{Inferring Sociodemographics on Reddit} \label{subsec:incorporating-socdem}

We estimate users' positions along latent social dimensions by leveraging the subreddit embedding scores introduced by \citet{wallerQuantifyingSocialOrganization2021}.
These embeddings represent \num{10060} popular subreddits from the 2005--2018 period as $d$-dimensional vectors, positioning communities with similar user bases close together in the embedding space.

\subsubsection{Social dimension scores.}

Social dimensions (e.g., age, gender, affluence, and partisanship) are derived by identifying pairs of seed subreddits that are strongly differentiated along a target construct but otherwise comparable---for example, r/teenagers versus r/RedditForGrownups for age.
For each construct, the method selects the top-$k$ seed pairs whose vector differences most closely align with the seed difference, then averages these differences to obtain a stable direction in embedding space (see \citet{wallerQuantifyingSocialOrganization2021} for full details and the complete seed list).
Each subreddit is then assigned a score on a given dimension by projecting its embedding onto this direction vector: subreddits closer to one pole (e.g.,
``younger'' or ``older'') receive more extreme values, while those equally related to both poles lie near zero (see \Cref{fig:score-distributions} for the complete score distributions).

\subsubsection{User-level scores.} 

Following an approach similar to \citet{montiEvidenceDemographicRather2023}, we map subreddit-level scores to individual users via their posting activity.
For user $u$ and dimension $d$, we define the user score as the post-weighted average of the subreddits in which the user participates:
\begin{equation}
F_{u}^{d} = \frac{\sum_{s}N_{u,s} F^{d}_{s}}{\sum_{s}N_{u,s}},
\end{equation}

where $N_{u,s}$ denotes the number of posts by user $u$ in subreddit $s$, and $F^{d}_{s}$ is the subreddit’s score on dimension $d$.

We compute user-level scores for age, gender, affluence, and partisanship for all users active across the 89 157 subreddits in our dataset.
Because the embeddings of \citet{wallerQuantifyingSocialOrganization2021} are defined over an independent set of \num{10060} subreddits, we restrict analysis to the \num{8296} that remain active within our observation window.
To reduce noise, we further consider only users who have posted at least five times in subreddits for which embedding scores are available.

It is important to note that the resulting user scores capture
\emph{associations} with social constructs rather than directly measured individual attributes.
For example, a user's position on the partisanship dimension reflects alignment with left- or right-leaning discourse on Reddit, not a measurement of actual political ideology.
We validated the age dimension scores against self-reported ages from the same users on Reddit: users scoring toward the ``older'' pole tend to be above 25 years old, while those scoring toward the ``younger'' pole tend to be below 20 years old (see \Cref{fig:age-comparison}).

\subsection{Stratified Attention Flow Graph}\label{subsec:strat_flow_graph}

We construct the Stratified Attention Flow Graph (S-AFG) as a
sociodemographically conditioned extension of the Attention Flow Graph (AFG) developed by \citet{rolloCommunitiesGatewaysBridges2022}.
The S-AFG captures how flows of user attention between subreddits differ across latent social dimensions such as age, gender, affluence, and partisanship.
We proceed in two steps: we first build the underlying AFG that serves as the structural backbone, and then
introduce a stratification procedure that conditions these flows on users' sociodemographic profiles.

\subsubsection{Building the AFG.}
The AFG models how user attention shifts between subreddits over time.
It is a weighted, directed network in which an edge from subreddit $i$ to subreddit $j$ represents the average migration of user attention between the two communities (see \Cref{si:network-construction} for details).
This representation encodes Reddit's topical structure through observed transitions in user activity.

Because the resulting AFG is extremely dense (over 2.2M edges among 65k nodes), we apply a disparity filter \citep{serranoExtractingMultiscaleBackbone2009} to remove statistically insignificant edges under a local null model of weight allocation (see \Cref{si:network-construction}).
Using a significance threshold of $\alpha = 0.01$, this yields a sparse backbone of \num{779021} directed edges over \num{62139} nodes (giant component).
As shown in \Cref{fig:weights_distribution}, this procedure preserves the heavy-tailed weight distribution while substantially reducing network density, enabling downstream analysis without distorting the global structure of flows.

\subsubsection{Incorporating Sociodemographics.}

To incorporate sociodemographic structure into attention flows, we extend the
AFG by weighting each user's contribution to the network by their score on a given social dimension.
Let $F_u^{d}$ denote the score of user $u$ on dimension $d$.
We construct a dimension-specific flow network as:
\begin{equation}
\mathbf{F}_{d}^{(t)} = \sum_{u\in \mathcal{U}} \mathbf{F}^{(t,u)}  \cdot F_u^{d},
\end{equation}

where $\mathbf{F}^{(t,u)}$ represents the attention flow contributed by user $u$ at time $t$.
We then aggregate these flows over time to obtain the final Stratified Attention Flow Graph for each dimension:
\begin{equation}
\mathcal{F}_{d} = \frac{1}{T} \sum_{t=1}^{T}\mathbf{F}_{d}^{(t)}.
\end{equation}

To ensure that stratified flows reflect observed transitions in the original network, we reweight edges by the corresponding AFG weights and restrict analysis to edges present in the original graph.

Because user scores can be positive or negative along a dimension, $\mathcal{F}_{d}$ is a signed network.
We therefore decompose it into positive and negative components, $\mathcal{F}_d^{+}$ and $\mathcal{F}_d^{-}$, corresponding to users aligned with opposite ends of each social dimension.
For example, edges in $\mathcal{F}_{\text{age}}^{+}$ represent attention flows associated with users positioned toward the ``older'' end of the age dimension, while $\mathcal{F}_{\text{age}}^{-}$ captures flows associated with younger users.

The resulting S-AFG provides a principled framework for comparing attention pathways across sociodemographic groups.
In particular, it enables us to identify subreddits that act as \emph{gateways} or \emph{bridges} into politically and conspiratorially oriented communities, and to quantify how these entry and transition structures vary across social dimensions.

\subsection{Measures on the (S-)AFG}

This subsection describes four measures computed on the (S-)AFG:
(i) Community detection, used to isolate communities of interest;
(ii) Sociodemographic profiling of community entrants;
(iii) Gateway and bridge identification using Personalized PageRank;
(iv) Shared-URL analysis to characterise prior information environments.

\subsubsection{Community Detection.} 

We detect topical communities on the filtered AFG using the Infomap algorithm \citep{rosvallInformationtheoreticFrameworkResolving2007, rosvallMapsRandomWalks2008}.
We prefer Infomap over inference via stochastic block models (SBMs) \citep{peixotoNonparametricBayesianInference2017} due to its computational scalability, as inferential methods are prohibitive at this scale.
This yields \num{4083} communities with a power-law-like size distribution (see \Cref{fig:community_size}).

We use these communities to isolate politically and conspiratorially oriented fora.
Building on the communities identified by \citet{rolloCommunitiesGatewaysBridges2022}, we select five communities of interest spanning conspiratorial, alt-right, radical left, and environmental topics.
For each, we select a representative seed subreddit and retrieve the detected community containing that seed (see \Cref{tab:afg_summary} for the seed subreddits and resulting community statistics).
These communities serve as the structural backbone for all subsequent analyses, anchoring the stratified flow measures and defining the units over which sociodemographic differences in attention are quantified.

\subsubsection{Sociodemographics of community entrants.}
We quantify the sociodemographic profiles of users \emph{prior to} their entry into each detected community.
For each community, we compute user-level scores using only posts made before a user's first participation in that community, excluding any activity within subreddits belonging to the community itself.
This procedure ensures that measurements are not influenced by exposure to the target community and instead reflect users' prior positioning in the broader attention network.
The resulting measures allow us to compare which sociodemographic groups are more likely to enter different types of communities, with a particular focus on political and conspiratorial fora.

\subsubsection{Gateways and Bridges.}
To characterise how users enter and transition into communities, we identify \emph{gateway} and \emph{bridge} subreddits using the S-AFG.
Following \citet{rolloCommunitiesGatewaysBridges2022}, we model Reddit users as discrete-time random walkers on the weighted adjacency matrix of the S-AFG, where transition probabilities are determined by edge weights.
We then compute Personalized PageRank (PPR) scores \citep{pagePagerankcitation1999, gleichPageRankWeb2015} with restart probabilities centred on a target community.
A gateway node (\emph{to} a community $X$) is a subreddit with a high PPR-derived probability of being an entry point for users arriving from outside $X$.
To identify \emph{bridges}, we compute PPR on the transpose of the S-AFG, effectively reversing edge directions so that highly ranked nodes correspond to subreddits lying on many incoming pathways
toward the target community.

Applying these definitions to the positive and negative components of the S-AFG allows us to compare how gateway and bridge subreddits differ across sociodemographic groups.
We quantify similarities and differences between these rankings using the weighted Kendall's $\tau$ correlation coefficient~\citep{vignaWeightedCorrelationIndex2015}.
Unlike standard Kendall $\tau$, the weighted variant assigns greater importance to highly ranked elements, making it particularly suitable for comparing top-ranked gateway and bridge subreddits while reducing noise from lower-ranked nodes.
Following \citet{vignaWeightedCorrelationIndex2015}, we adopt a hyperbolic weighting scheme that emphasises agreement among the most important subreddits in each ranking.
 
\subsubsection{Shared URLs Prior to Community Entry.}
To characterise users' information consumption patterns before entering a community, we extract the external URLs shared by users prior to their first participation in a given community, applying the same pre-entry restriction used for sociodemographic scores.
Specifically, for each target community, we consider only URLs shared in posts made before a user's first interaction with that community, excluding activity within subreddits belonging to the community itself.
This construction isolates users' prior information environments and allows us to examine whether distinct communities are associated with different patterns of external content consumption.
In particular, these measures provide a complementary view of potential offline or cross-platform associations that may not be fully captured by Reddit interaction patterns alone.

\section{Data availability}

All data and code required to reproduce the analyses and plots presented in this paper are openly available.
The processed data underlying the analyses can be found at
\url{https://doi.org/10.5281/zenodo.21416900},
and the corresponding analysis code is available at
\url{https://doi.org/10.5281/zenodo.21416527}.

The raw, unprocessed data used to construct the Attention-Flow Graph---and, consequently, all subsequent analyses derived from it---are not included in these repositories, as they are publicly available through the Pushshift dataset~\citep{baumgartnerPushshiftRedditDataset2020}.

\section{Ethics statement}

This study uses publicly available data from Reddit, collected via the Pushshift dataset.
No direct interaction with human subjects was involved, and no personally identifiable information was collected or stored.
Sociodemographic attributes were not directly observed but inferred statistically at the aggregate level from users' posting behavior across subreddits.
Individual users are not identified or profiled in any reported result.
All analyses were conducted at the community and group level.
This research was determined to be exempt from IRB review under the category of research involving publicly available data with no possibility of identifying individual subjects.

\bibliographystyle{ACM-Reference-Format}
\bibliography{../reference}


\begin{thebibliography}{38}


\ifx \showCODEN    \undefined \def \showCODEN     #1{\unskip}     \fi
\ifx \showISBNx    \undefined \def \showISBNx     #1{\unskip}     \fi
\ifx \showISBNxiii \undefined \def \showISBNxiii  #1{\unskip}     \fi
\ifx \showISSN     \undefined \def \showISSN      #1{\unskip}     \fi
\ifx \showLCCN     \undefined \def \showLCCN      #1{\unskip}     \fi
\ifx \shownote     \undefined \def \shownote      #1{#1}          \fi
\ifx \showarticletitle \undefined \def \showarticletitle #1{#1}   \fi
\ifx \showURL      \undefined \def \showURL       {\relax}        \fi
\providecommand\bibfield[2]{#2}
\providecommand\bibinfo[2]{#2}
\providecommand\natexlab[1]{#1}
\providecommand\showeprint[2][]{arXiv:#2}

\bibitem[Ang et~al\mbox{.}(2020)]%
        {angIndividualismAdoptionClean2020}
\bibfield{author}{\bibinfo{person}{James~B. Ang}, \bibinfo{person}{Per~G.
  Fredriksson}, {and} \bibinfo{person}{Swati Sharma}.}
  \bibinfo{year}{2020}\natexlab{}.
\newblock \showarticletitle{Individualism and the Adoption of Clean Energy
  Technology}.
\newblock \bibinfo{journal}{\emph{Resource and Energy Economics}}
  \bibinfo{volume}{61} (\bibinfo{date}{Aug.} \bibinfo{year}{2020}),
  \bibinfo{pages}{101180}.
\newblock
\showISSN{0928-7655}
\href{https://doi.org/10.1016/j.reseneeco.2020.101180}{doi:\nolinkurl{10.1016/j.reseneeco.2020.101180}}


\bibitem[Baumgartner et~al\mbox{.}(2020)]%
        {baumgartnerPushshiftRedditDataset2020}
\bibfield{author}{\bibinfo{person}{Jason Baumgartner}, \bibinfo{person}{Savvas
  Zannettou}, \bibinfo{person}{Brian Keegan}, \bibinfo{person}{Megan Squire},
  {and} \bibinfo{person}{Jeremy Blackburn}.} \bibinfo{year}{2020}\natexlab{}.
\newblock \showarticletitle{The {{Pushshift Reddit Dataset}}}.
\newblock \bibinfo{journal}{\emph{Proceedings of the International AAAI
  Conference on Web and Social Media}}  \bibinfo{volume}{14}
  (\bibinfo{date}{May} \bibinfo{year}{2020}), \bibinfo{pages}{830--839}.
\newblock
\showISSN{2334-0770}
\href{https://doi.org/10.1609/icwsm.v14i1.7347}{doi:\nolinkurl{10.1609/icwsm.v14i1.7347}}


\bibitem[Bishop and Cushing(2008)]%
        {bishopBigSortWhy2008}
\bibfield{author}{\bibinfo{person}{Bill Bishop} {and}
  \bibinfo{person}{Robert~G. Cushing}.} \bibinfo{year}{2008}\natexlab{}.
\newblock \bibinfo{booktitle}{\emph{The {{Big Sort}}: {{Why}} the
  {{Clustering}} of {{Like-minded America}} Is {{Tearing Us Apart}}}}.
\newblock \bibinfo{publisher}{Houghton Mifflin Harcourt}.
\newblock


\bibitem[Boulianne(2020)]%
        {boulianneTwentyYearsDigital2020}
\bibfield{author}{\bibinfo{person}{Shelley Boulianne}.}
  \bibinfo{year}{2020}\natexlab{}.
\newblock \showarticletitle{Twenty {{Years}} of {{Digital Media Effects}} on
  {{Civic}} and {{Political Participation}}}.
\newblock \bibinfo{journal}{\emph{Communication Research}}
  \bibinfo{volume}{47}, \bibinfo{number}{7} (\bibinfo{date}{Oct.}
  \bibinfo{year}{2020}), \bibinfo{pages}{947--966}.
\newblock
\showISSN{0093-6502}
\href{https://doi.org/10.1177/0093650218808186}{doi:\nolinkurl{10.1177/0093650218808186}}


\bibitem[Carman et~al\mbox{.}(2024)]%
        {UnderstandingProclimateVoters}
\bibfield{author}{\bibinfo{person}{Jennifer Carman}, \bibinfo{person}{Matthew
  Ballew}, \bibinfo{person}{Marija Verner}, \bibinfo{person}{Seth Rosenthal},
  \bibinfo{person}{Jan Kotcher}, \bibinfo{person}{Edward Maibach}, {and}
  \bibinfo{person}{Anthony Leiserowitz}.} \bibinfo{year}{2024}\natexlab{}.
\newblock \bibinfo{title}{Understanding Pro-Climate Voters in the {{United
  States}}}.
\newblock
\urldef\tempurl%
\url{https://climatecommunication.yale.edu/publications/understanding-pro-climate-voters/}
\showURL{%
\tempurl}


\bibitem[Cho et~al\mbox{.}(2013)]%
        {choBeNotBe2013}
\bibfield{author}{\bibinfo{person}{Yoon-Na Cho}, \bibinfo{person}{Anastasia
  Thyroff}, \bibinfo{person}{Molly~I. Rapert}, \bibinfo{person}{Seong-Yeon
  Park}, {and} \bibinfo{person}{Hyun~Ju Lee}.} \bibinfo{year}{2013}\natexlab{}.
\newblock \showarticletitle{To Be or Not to Be Green: {{Exploring}}
  Individualism and Collectivism as Antecedents of Environmental Behavior}.
\newblock \bibinfo{journal}{\emph{Journal of Business Research}}
  \bibinfo{volume}{66}, \bibinfo{number}{8} (\bibinfo{date}{Aug.}
  \bibinfo{year}{2013}), \bibinfo{pages}{1052--1059}.
\newblock
\showISSN{0148-2963}
\href{https://doi.org/10.1016/j.jbusres.2012.08.020}{doi:\nolinkurl{10.1016/j.jbusres.2012.08.020}}


\bibitem[Cinelli et~al\mbox{.}(2021)]%
        {cinelliEchoChamberEffect2021a}
\bibfield{author}{\bibinfo{person}{Matteo Cinelli}, \bibinfo{person}{Gianmarco
  De~Francisci~Morales}, \bibinfo{person}{Alessandro Galeazzi},
  \bibinfo{person}{Walter Quattrociocchi}, {and} \bibinfo{person}{Michele
  Starnini}.} \bibinfo{year}{2021}\natexlab{}.
\newblock \showarticletitle{The Echo Chamber Effect on Social Media}.
\newblock \bibinfo{journal}{\emph{Proceedings of the National Academy of
  Sciences}} \bibinfo{volume}{118}, \bibinfo{number}{9} (\bibinfo{date}{March}
  \bibinfo{year}{2021}), \bibinfo{pages}{e2023301118}.
\newblock
\href{https://doi.org/10.1073/pnas.2023301118}{doi:\nolinkurl{10.1073/pnas.2023301118}}


\bibitem[Cinus et~al\mbox{.}(2025)]%
        {cinusUncoveringSociodemographicFabric2025}
\bibfield{author}{\bibinfo{person}{Federico Cinus}, \bibinfo{person}{Corrado
  Monti}, \bibinfo{person}{Paolo Bajardi}, {and} \bibinfo{person}{Gianmarco
  De~Francisci Morales}.} \bibinfo{year}{2025}\natexlab{}.
\newblock \bibinfo{title}{Uncovering the {{Sociodemographic Fabric}} of
  {{Reddit}}}.
\newblock
\href{https://doi.org/10.48550/arXiv.2502.05049}{doi:\nolinkurl{10.48550/arXiv.2502.05049}}
\showeprint{2502.05049}~[cs.SI]


\bibitem[Datta and Adar(2019)]%
        {dattaExtractingInterCommunityConflicts2019}
\bibfield{author}{\bibinfo{person}{Srayan Datta} {and} \bibinfo{person}{Eytan
  Adar}.} \bibinfo{year}{2019}\natexlab{}.
\newblock \showarticletitle{Extracting {{Inter-Community Conflicts}} in
  {{Reddit}}}.
\newblock \bibinfo{journal}{\emph{Proceedings of the International AAAI
  Conference on Web and Social Media}}  \bibinfo{volume}{13}
  (\bibinfo{date}{July} \bibinfo{year}{2019}), \bibinfo{pages}{146--157}.
\newblock
\showISSN{2334-0770}
\href{https://doi.org/10.1609/icwsm.v13i01.3217}{doi:\nolinkurl{10.1609/icwsm.v13i01.3217}}


\bibitem[Del~Vicario et~al\mbox{.}(2016)]%
        {delvicarioSpreadingMisinformationOnline2016}
\bibfield{author}{\bibinfo{person}{Michela Del~Vicario},
  \bibinfo{person}{Alessandro Bessi}, \bibinfo{person}{Fabiana Zollo},
  \bibinfo{person}{Fabio Petroni}, \bibinfo{person}{Antonio Scala},
  \bibinfo{person}{Guido Caldarelli}, \bibinfo{person}{H.~Eugene Stanley},
  {and} \bibinfo{person}{Walter Quattrociocchi}.}
  \bibinfo{year}{2016}\natexlab{}.
\newblock \showarticletitle{The Spreading of Misinformation Online}.
\newblock \bibinfo{journal}{\emph{Proceedings of the National Academy of
  Sciences}} \bibinfo{volume}{113}, \bibinfo{number}{3} (\bibinfo{date}{Jan.}
  \bibinfo{year}{2016}), \bibinfo{pages}{554--559}.
\newblock
\href{https://doi.org/10.1073/pnas.1517441113}{doi:\nolinkurl{10.1073/pnas.1517441113}}


\bibitem[Douglas et~al\mbox{.}(2019)]%
        {douglasUnderstandingConspiracyTheories2019}
\bibfield{author}{\bibinfo{person}{Karen~M. Douglas},
  \bibinfo{person}{Joseph~E. Uscinski}, \bibinfo{person}{Robbie~M. Sutton},
  \bibinfo{person}{Aleksandra Cichocka}, \bibinfo{person}{Turkay Nefes},
  \bibinfo{person}{Chee~Siang Ang}, {and} \bibinfo{person}{Farzin Deravi}.}
  \bibinfo{year}{2019}\natexlab{}.
\newblock \showarticletitle{Understanding {{Conspiracy Theories}}}.
\newblock \bibinfo{journal}{\emph{Political Psychology}} \bibinfo{volume}{40},
  \bibinfo{number}{S1} (\bibinfo{year}{2019}), \bibinfo{pages}{3--35}.
\newblock
\showISSN{1467-9221}
\href{https://doi.org/10.1111/pops.12568}{doi:\nolinkurl{10.1111/pops.12568}}


\bibitem[Engel et~al\mbox{.}(2022)]%
        {engelCharacterizingRedditParticipation2022}
\bibfield{author}{\bibinfo{person}{Kristen Engel}, \bibinfo{person}{Yiqing
  Hua}, \bibinfo{person}{Taixiang Zeng}, {and} \bibinfo{person}{Mor Naaman}.}
  \bibinfo{year}{2022}\natexlab{}.
\newblock \showarticletitle{Characterizing {{Reddit Participation}} of {{Users
  Who Engage}} in the {{QAnon Conspiracy Theories}}}.
\newblock \bibinfo{journal}{\emph{Proceedings of the ACM on Human-Computer
  Interaction}} \bibinfo{volume}{6}, \bibinfo{number}{CSCW1}
  (\bibinfo{date}{April} \bibinfo{year}{2022}), \bibinfo{pages}{53:1--53:22}.
\newblock
\href{https://doi.org/10.1145/3512900}{doi:\nolinkurl{10.1145/3512900}}


\bibitem[{Fage-Butler} et~al\mbox{.}(2022)]%
        {fage-butlerPublicTrustMistrust2022}
\bibfield{author}{\bibinfo{person}{Antoinette {Fage-Butler}},
  \bibinfo{person}{Loni Ledderer}, {and} \bibinfo{person}{Kristian~Hvidtfelt
  Nielsen}.} \bibinfo{year}{2022}\natexlab{}.
\newblock \showarticletitle{Public Trust and Mistrust of Climate Science: {{A}}
  Meta-Narrative Review}.
\newblock \bibinfo{journal}{\emph{Public Understanding of Science}}
  \bibinfo{volume}{31}, \bibinfo{number}{7} (\bibinfo{date}{Oct.}
  \bibinfo{year}{2022}), \bibinfo{pages}{832--846}.
\newblock
\showISSN{0963-6625}
\href{https://doi.org/10.1177/09636625221110028}{doi:\nolinkurl{10.1177/09636625221110028}}


\bibitem[Gleich(2015)]%
        {gleichPageRankWeb2015}
\bibfield{author}{\bibinfo{person}{David~F. Gleich}.}
  \bibinfo{year}{2015}\natexlab{}.
\newblock \showarticletitle{{{PageRank Beyond}} the {{Web}}}.
\newblock \bibinfo{journal}{\emph{SIAM Rev.}} \bibinfo{volume}{57},
  \bibinfo{number}{3} (\bibinfo{date}{Jan.} \bibinfo{year}{2015}),
  \bibinfo{pages}{321--363}.
\newblock
\showISSN{0036-1445}
\href{https://doi.org/10.1137/140976649}{doi:\nolinkurl{10.1137/140976649}}


\bibitem[Klandermans(2004)]%
        {klandermansDemandSupplyParticipation2004}
\bibfield{author}{\bibinfo{person}{Bert Klandermans}.}
  \bibinfo{year}{2004}\natexlab{}.
\newblock \bibinfo{booktitle}{\emph{The {{Demand}} and {{Supply}} of
  {{Participation}}: {{Social-Psychological Correlates}} of {{Participation}}
  in {{Social Movements}}}}.
\newblock \bibinfo{publisher}{John Wiley \& Sons, Ltd}, Chapter~16,
  \bibinfo{pages}{360--379}.
\newblock
\href{https://doi.org/10.1002/9780470999103.ch16}{doi:\nolinkurl{10.1002/9780470999103.ch16}}


\bibitem[Lenti et~al\mbox{.}(2025)]%
        {lentiCausalModelingClimate2025}
\bibfield{author}{\bibinfo{person}{Jacopo Lenti}, \bibinfo{person}{Luca~Maria
  Aiello}, \bibinfo{person}{Corrado Monti}, {and} \bibinfo{person}{Gianmarco
  De~Francisci Morales}.} \bibinfo{year}{2025}\natexlab{}.
\newblock \showarticletitle{Causal {{Modeling}} of {{Climate Activism}} on
  {{Reddit}}}. In \bibinfo{booktitle}{\emph{Proceedings of the {{ACM}} on {{Web
  Conference}} 2025}} \emph{(\bibinfo{series}{{{WWW}} '25})}.
  \bibinfo{publisher}{Association for Computing Machinery},
  \bibinfo{address}{New York, NY, USA}, \bibinfo{pages}{590--600}.
\newblock
\showISBNx{979-8-4007-1274-6}
\href{https://doi.org/10.1145/3696410.3714684}{doi:\nolinkurl{10.1145/3696410.3714684}}


\bibitem[Mason(2018)]%
        {masonUncivilAgreement2018}
\bibfield{author}{\bibinfo{person}{Lilliana Mason}.}
  \bibinfo{year}{2018}\natexlab{}.
\newblock \bibinfo{booktitle}{\emph{Uncivil {{Agreement: How Politics Became
  Our Identity}}}}.
\newblock \bibinfo{publisher}{University of Chicago Press}.
\newblock
\showISBNx{978-978-022-652-7}


\bibitem[Monti et~al\mbox{.}(2023)]%
        {montiEvidenceDemographicRather2023}
\bibfield{author}{\bibinfo{person}{Corrado Monti}, \bibinfo{person}{Jacopo
  D'Ignazi}, \bibinfo{person}{Michele Starnini}, {and}
  \bibinfo{person}{Gianmarco De~Francisci~Morales}.}
  \bibinfo{year}{2023}\natexlab{}.
\newblock \showarticletitle{Evidence of {{Demographic}} Rather than
  {{Ideological Segregation}} in {{News Discussion}} on {{Reddit}}}. In
  \bibinfo{booktitle}{\emph{Proceedings of the {{ACM Web Conference}} 2023}}
  (Austin, TX, USA) \emph{(\bibinfo{series}{{{WWW}} '23})}.
  \bibinfo{publisher}{Association for Computing Machinery},
  \bibinfo{address}{New York, NY, USA}, \bibinfo{pages}{2777--2786}.
\newblock
\href{https://doi.org/10.1145/3543507.3583468}{doi:\nolinkurl{10.1145/3543507.3583468}}


\bibitem[Neas et~al\mbox{.}(2022)]%
        {neasYoungPeoplesClimate2022}
\bibfield{author}{\bibinfo{person}{Sally Neas}, \bibinfo{person}{Ann Ward},
  {and} \bibinfo{person}{Benjamin Bowman}.} \bibinfo{year}{2022}\natexlab{}.
\newblock \showarticletitle{Young People's Climate Activism: {{A}} Review of
  the Literature}.
\newblock \bibinfo{journal}{\emph{Frontiers in Political Science}}
  \bibinfo{volume}{4} (\bibinfo{date}{Aug.} \bibinfo{year}{2022}).
\newblock
\showISSN{2673-3145}
\href{https://doi.org/10.3389/fpos.2022.940876}{doi:\nolinkurl{10.3389/fpos.2022.940876}}


\bibitem[Page et~al\mbox{.}(1999)]%
        {pagePagerankcitation1999}
\bibfield{author}{\bibinfo{person}{Lawrence Page}, \bibinfo{person}{Sergey
  Brin}, \bibinfo{person}{Rajeev Motwani}, {and} \bibinfo{person}{Terry
  Winograd}.} \bibinfo{year}{1999}\natexlab{}.
\newblock \showarticletitle{The {PageRank Citation Ranking: Bringing Order to
  the Web}}. In \bibinfo{booktitle}{\emph{Proceedings of the 7th
  {{International Conference}} on {{World Wide Web}}}}.
\newblock


\bibitem[Pariser(2011)]%
        {pariserFilterBubbleWhat2011}
\bibfield{author}{\bibinfo{person}{Eli Pariser}.}
  \bibinfo{year}{2011}\natexlab{}.
\newblock \bibinfo{booktitle}{\emph{The {{Filter Bubble}}: {{What The Internet
  Is Hiding From You}}}}.
\newblock \bibinfo{publisher}{Penguin UK}.
\newblock
\showISBNx{978-0-14-196992-3}


\bibitem[Peixoto(2017)]%
        {peixotoNonparametricBayesianInference2017}
\bibfield{author}{\bibinfo{person}{Tiago~P. Peixoto}.}
  \bibinfo{year}{2017}\natexlab{}.
\newblock \showarticletitle{Nonparametric {{Bayesian}} Inference of the
  Microcanonical Stochastic Block Model}.
\newblock \bibinfo{journal}{\emph{Physical Review E}} \bibinfo{volume}{95},
  \bibinfo{number}{1} (\bibinfo{date}{Jan.} \bibinfo{year}{2017}),
  \bibinfo{pages}{012317}.
\newblock
\href{https://doi.org/10.1103/PhysRevE.95.012317}{doi:\nolinkurl{10.1103/PhysRevE.95.012317}}


\bibitem[{Pew Research Center}(2020)]%
        {pew5FactsQAnon2020}
\bibfield{author}{\bibinfo{person}{{Pew Research Center}}.}
  \bibinfo{year}{2020}\natexlab{}.
\newblock \bibinfo{title}{5 Facts about the {{QAnon}} Conspiracy Theories}.
\newblock
\urldef\tempurl%
\url{https://www.pewresearch.org/short-reads/2020/11/16/5-facts-about-the-qanon-conspiracy-theories/}
\showURL{%
\tempurl}


\bibitem[Polletta and Jasper(2001)]%
        {pollettaCollectiveIdentitySocial2001}
\bibfield{author}{\bibinfo{person}{Francesca Polletta} {and}
  \bibinfo{person}{James~M. Jasper}.} \bibinfo{year}{2001}\natexlab{}.
\newblock \showarticletitle{Collective {{Identity}} and {{Social Movements}}}.
\newblock \bibinfo{journal}{\emph{Annual Review of Sociology}}
  \bibinfo{volume}{27} (\bibinfo{year}{2001}), \bibinfo{pages}{283--305}.
\newblock
\showISSN{0360-0572}
\showeprint[jstor]{2678623}


\bibitem[Rollo et~al\mbox{.}(2022)]%
        {rolloCommunitiesGatewaysBridges2022}
\bibfield{author}{\bibinfo{person}{Cesare Rollo}, \bibinfo{person}{Gianmarco
  De~Francisci~Morales}, \bibinfo{person}{Corrado Monti}, {and}
  \bibinfo{person}{Andr{\'e} Panisson}.} \bibinfo{year}{2022}\natexlab{}.
\newblock \showarticletitle{Communities, {{Gateways}}, and~{{Bridges}}:
  {{Measuring Attention Flow}} in~the~{{Reddit Political Sphere}}}. In
  \bibinfo{booktitle}{\emph{Social {{Informatics}}: 13th {{International
  Conference}}, {{SocInfo}} 2022, {{Glasgow}}, {{UK}}, {{October}} 19--21,
  2022, {{Proceedings}}}}. \bibinfo{pages}{3--19}.
\newblock
\href{https://doi.org/10.1007/978-3-031-19097-1_1}{doi:\nolinkurl{10.1007/978-3-031-19097-1_1}}


\bibitem[Rosvall and Bergstrom(2007)]%
        {rosvallInformationtheoreticFrameworkResolving2007}
\bibfield{author}{\bibinfo{person}{Martin Rosvall} {and}
  \bibinfo{person}{Carl~T. Bergstrom}.} \bibinfo{year}{2007}\natexlab{}.
\newblock \showarticletitle{An Information-Theoretic Framework for Resolving
  Community Structure in Complex Networks}.
\newblock \bibinfo{journal}{\emph{Proceedings of the National Academy of
  Sciences}} \bibinfo{volume}{104}, \bibinfo{number}{18} (\bibinfo{date}{May}
  \bibinfo{year}{2007}), \bibinfo{pages}{7327--7331}.
\newblock
\href{https://doi.org/10.1073/pnas.0611034104}{doi:\nolinkurl{10.1073/pnas.0611034104}}


\bibitem[Rosvall and Bergstrom(2008)]%
        {rosvallMapsRandomWalks2008}
\bibfield{author}{\bibinfo{person}{Martin Rosvall} {and}
  \bibinfo{person}{Carl~T. Bergstrom}.} \bibinfo{year}{2008}\natexlab{}.
\newblock \showarticletitle{Maps of Random Walks on Complex Networks Reveal
  Community Structure}.
\newblock \bibinfo{journal}{\emph{Proceedings of the National Academy of
  Sciences}} \bibinfo{volume}{105}, \bibinfo{number}{4} (\bibinfo{date}{Jan.}
  \bibinfo{year}{2008}), \bibinfo{pages}{1118--1123}.
\newblock
\href{https://doi.org/10.1073/pnas.0706851105}{doi:\nolinkurl{10.1073/pnas.0706851105}}


\bibitem[Schlosberg and Collins(2014)]%
        {schlosbergEnvironmentalClimateJustice2014}
\bibfield{author}{\bibinfo{person}{David Schlosberg} {and}
  \bibinfo{person}{Lisette~B. Collins}.} \bibinfo{year}{2014}\natexlab{}.
\newblock \showarticletitle{From Environmental to Climate Justice: Climate
  Change and the Discourse of Environmental Justice}.
\newblock \bibinfo{journal}{\emph{WIREs Climate Change}} \bibinfo{volume}{5},
  \bibinfo{number}{3} (\bibinfo{year}{2014}), \bibinfo{pages}{359--374}.
\newblock
\showISSN{1757-7799}
\href{https://doi.org/10.1002/wcc.275}{doi:\nolinkurl{10.1002/wcc.275}}


\bibitem[Serrano et~al\mbox{.}(2009)]%
        {serranoExtractingMultiscaleBackbone2009}
\bibfield{author}{\bibinfo{person}{M.~{\'A}ngeles Serrano},
  \bibinfo{person}{Mari{\'a}n Bogu{\~n}{\'a}}, {and}
  \bibinfo{person}{Alessandro Vespignani}.} \bibinfo{year}{2009}\natexlab{}.
\newblock \showarticletitle{Extracting the Multiscale Backbone of Complex
  Weighted Networks}.
\newblock \bibinfo{journal}{\emph{Proceedings of the National Academy of
  Sciences}} \bibinfo{volume}{106}, \bibinfo{number}{16} (\bibinfo{date}{April}
  \bibinfo{year}{2009}), \bibinfo{pages}{6483--6488}.
\newblock
\href{https://doi.org/10.1073/pnas.0808904106}{doi:\nolinkurl{10.1073/pnas.0808904106}}


\bibitem[Sharma et~al\mbox{.}(2022)]%
        {sharmaCharacterizingOnlineEngagement2022}
\bibfield{author}{\bibinfo{person}{Karishma Sharma}, \bibinfo{person}{Emilio
  Ferrara}, {and} \bibinfo{person}{Yan Liu}.} \bibinfo{year}{2022}\natexlab{}.
\newblock \showarticletitle{Characterizing {{Online Engagement}} with
  {{Disinformation}} and {{Conspiracies}} in the 2020 {{U}}.{{S}}.
  {{Presidential Election}}}.
\newblock \bibinfo{journal}{\emph{Proceedings of the International AAAI
  Conference on Web and Social Media}}  \bibinfo{volume}{16}
  (\bibinfo{date}{May} \bibinfo{year}{2022}), \bibinfo{pages}{908--919}.
\newblock
\showISSN{2334-0770}
\href{https://doi.org/10.1609/icwsm.v16i1.19345}{doi:\nolinkurl{10.1609/icwsm.v16i1.19345}}


\bibitem[Sternisko et~al\mbox{.}(2020)]%
        {sterniskoDarkSideSocial2020}
\bibfield{author}{\bibinfo{person}{Anni Sternisko}, \bibinfo{person}{Aleksandra
  Cichocka}, {and} \bibinfo{person}{Jay~J Van~Bavel}.}
  \bibinfo{year}{2020}\natexlab{}.
\newblock \showarticletitle{The Dark Side of Social Movements: Social Identity,
  Non-Conformity, and the Lure of Conspiracy Theories}.
\newblock \bibinfo{journal}{\emph{Current Opinion in Psychology}}
  \bibinfo{volume}{35} (\bibinfo{date}{Oct.} \bibinfo{year}{2020}),
  \bibinfo{pages}{1--6}.
\newblock
\showISSN{2352-250X}
\href{https://doi.org/10.1016/j.copsyc.2020.02.007}{doi:\nolinkurl{10.1016/j.copsyc.2020.02.007}}


\bibitem[Sutton and Douglas(2020)]%
        {suttonConspiracyTheoriesConspiracy2020}
\bibfield{author}{\bibinfo{person}{Robbie~M Sutton} {and}
  \bibinfo{person}{Karen~M Douglas}.} \bibinfo{year}{2020}\natexlab{}.
\newblock \showarticletitle{Conspiracy Theories and the Conspiracy Mindset:
  Implications for Political Ideology}.
\newblock \bibinfo{journal}{\emph{Current Opinion in Behavioral Sciences}}
  \bibinfo{volume}{34} (\bibinfo{date}{Aug.} \bibinfo{year}{2020}),
  \bibinfo{pages}{118--122}.
\newblock
\showISSN{2352-1546}
\href{https://doi.org/10.1016/j.cobeha.2020.02.015}{doi:\nolinkurl{10.1016/j.cobeha.2020.02.015}}


\bibitem[Suzuki et~al\mbox{.}(2022)]%
        {YouthConcernClimate2022}
\bibfield{author}{\bibinfo{person}{Sara Suzuki}, \bibinfo{person}{Alberto
  Medina}, {and} \bibinfo{person}{Peter de Guzman}.}
  \bibinfo{year}{2022}\natexlab{}.
\newblock \bibinfo{title}{Youth {{Concern About Climate Change Drives Civic
  Engagement}}}.
\newblock \bibinfo{howpublished}{Center for Information \& Research on Civic
  Learning and Engagement}.
\newblock
\urldef\tempurl%
\url{https://circle.tufts.edu/latest-research/youth-concern-about-climate-change-drives-civic-engagement}
\showURL{%
\tempurl}


\bibitem[Uscinski et~al\mbox{.}(2022)]%
        {uscinskiPsychologicalPoliticalCorrelates2022}
\bibfield{author}{\bibinfo{person}{Joseph Uscinski}, \bibinfo{person}{Adam
  Enders}, \bibinfo{person}{Amanda Diekman}, \bibinfo{person}{John Funchion},
  \bibinfo{person}{Casey Klofstad}, \bibinfo{person}{Sandra Kuebler},
  \bibinfo{person}{Manohar Murthi}, \bibinfo{person}{Kamal Premaratne},
  \bibinfo{person}{Michelle Seelig}, \bibinfo{person}{Daniel Verdear}, {and}
  \bibinfo{person}{Stefan Wuchty}.} \bibinfo{year}{2022}\natexlab{}.
\newblock \showarticletitle{The Psychological and Political Correlates of
  Conspiracy Theory Beliefs}.
\newblock \bibinfo{journal}{\emph{Scientific Reports}} \bibinfo{volume}{12},
  \bibinfo{number}{1} (\bibinfo{date}{Dec.} \bibinfo{year}{2022}),
  \bibinfo{pages}{21672}.
\newblock
\showISSN{2045-2322}
\href{https://doi.org/10.1038/s41598-022-25617-0}{doi:\nolinkurl{10.1038/s41598-022-25617-0}}


\bibitem[{van Stekelenburg} and Klandermans(2013)]%
        {vanstekelenburgSocialPsychologyProtest2013}
\bibfield{author}{\bibinfo{person}{Jacquelien {van Stekelenburg}} {and}
  \bibinfo{person}{Bert Klandermans}.} \bibinfo{year}{2013}\natexlab{}.
\newblock \showarticletitle{The Social Psychology of Protest}.
\newblock \bibinfo{journal}{\emph{Current Sociology}} \bibinfo{volume}{61},
  \bibinfo{number}{5-6} (\bibinfo{date}{Sept.} \bibinfo{year}{2013}),
  \bibinfo{pages}{886--905}.
\newblock
\showISSN{0011-3921}
\href{https://doi.org/10.1177/0011392113479314}{doi:\nolinkurl{10.1177/0011392113479314}}


\bibitem[Vigna(2015)]%
        {vignaWeightedCorrelationIndex2015}
\bibfield{author}{\bibinfo{person}{Sebastiano Vigna}.}
  \bibinfo{year}{2015}\natexlab{}.
\newblock \showarticletitle{A {{Weighted Correlation Index}} for {{Rankings}}
  with {{Ties}}}. In \bibinfo{booktitle}{\emph{Proceedings of the 24th
  {{International Conference}} on {{World Wide Web}}}}
  \emph{(\bibinfo{series}{{{WWW}} '15})}. \bibinfo{pages}{1166--1176}.
\newblock
\href{https://doi.org/10.1145/2736277.2741088}{doi:\nolinkurl{10.1145/2736277.2741088}}


\bibitem[Wahlstr{\"o}m et~al\mbox{.}(2020)]%
        {wahlstromSurveysParticipantsFridays2020}
\bibfield{author}{\bibinfo{person}{Mattias Wahlstr{\"o}m},
  \bibinfo{person}{Joost {de Moor}}, \bibinfo{person}{Katrin Uba},
  \bibinfo{person}{Magnus Wennerhag}, \bibinfo{person}{Michiel De~Vydt},
  \bibinfo{person}{Paul Almeida}, \bibinfo{person}{Anja Baukloh},
  \bibinfo{person}{Niccol{\`o} Bertuzzi}, \bibinfo{person}{Daniela Chironi},
  {and} \bibinfo{person}{Brendan Churchill}.} \bibinfo{year}{2020}\natexlab{}.
\newblock \bibinfo{title}{Surveys of Participants in {{Fridays For Future}}
  Climate Protests on 20-28 {{September}}, 2019, in 19 Cities around the
  World}.
\newblock


\bibitem[Waller and Anderson(2021)]%
        {wallerQuantifyingSocialOrganization2021}
\bibfield{author}{\bibinfo{person}{Isaac Waller} {and} \bibinfo{person}{Ashton
  Anderson}.} \bibinfo{year}{2021}\natexlab{}.
\newblock \showarticletitle{Quantifying Social Organization and Political
  Polarization in Online Platforms}.
\newblock \bibinfo{journal}{\emph{Nature}} \bibinfo{volume}{600},
  \bibinfo{number}{7888} (\bibinfo{date}{Dec.} \bibinfo{year}{2021}),
  \bibinfo{pages}{264--268}.
\newblock
\showISSN{1476-4687}
\href{https://doi.org/10.1038/s41586-021-04167-x}{doi:\nolinkurl{10.1038/s41586-021-04167-x}}


\end{thebibliography}

\clearpage
\appendix

\renewcommand{\thesection}{S\arabic{section}}
\renewcommand{\thefigure}{S\arabic{figure}}
\renewcommand{\thetable}{S\arabic{table}}
\renewcommand{\theequation}{S\arabic{equation}}

\setcounter{section}{0}
\setcounter{figure}{0}
\setcounter{table}{0}
\setcounter{equation}{0}

\section*{Supplementary Material}
\addcontentsline{toc}{section}{Supplementary Material}

This Supplementary Material provides additional methodological details, analyses, and robustness checks that support the main findings of this study. It includes extended descriptions of the data, and procedures, as well as supplementary results that further validate the reported effects.

\section{Network construction}
\label{si:network-construction}

Here we provide the theoretical background for building the AFG as defined by \citet{rolloCommunitiesGatewaysBridges2022}, together with descriptive statistics of the final network and robustness checks of the network construction methodology.

\subsection{Attention Flow Graph} \label{sm:AFG}

The Attention Flow Graph (AFG), as developed by \citet{rolloCommunitiesGatewaysBridges2022}, is a network that summarizes how users shift their attention from one subreddit to another over time, capturing Reddit's topical community structure.
Formally, the AFG is a weighted and directed network, where an edge between subreddits represents how much of a user's attention migrated from a subreddit $i$ to a subreddit $j$ on average for a given time period.
Given a timestep $t$ (a given month), a set $\mathcal{U}$ of users and a set $\mathcal{S}$ of subreddits, the AFG for the Reddit environment is constructed by performing the following steps:

\begin{enumerate}
\item We obtain $c_{u,s}^{(t)}$, the number of interactions of user $u\in \mathcal{U}$ with subreddit $s\in \mathcal{S}$ at time $t$.
We define $\mathbf{b}_u^t$ as the \emph{attention vector} of user $u$ at time $t$, where each entry of the vector is given by the normalized value of $c_{u,s}^{(t)}$.

\item For each user $u$ and subreddit $s$, we quantify how much of the user’s attention changes between $t$ and $t' = t-1$ by obtaining:
\begin{equation*}
\Delta \mathbf{b}_{u,s}^t = 
\begin{cases}
\mathbf{b}_{u,s}^t - \mathbf{b}_{u,s}^{t'}, \quad& \text{if }\mathbf{b}_{u,s}^t \cdot \mathbf{b}_{u,s}^{t'} = 0\\
0, \quad& \text{otherwise}
\end{cases}.
\end{equation*}

We distinguish between the positive flow vector $\Delta^{+} \mathbf{b}_{u}^t$, that highlights adopted subreddits, and the negative flow vector $\Delta^{-} \mathbf{b}_{u}^t$, that highlights abandoned subreddits.

\item The \emph{attention flow} of a user $u$ at time $t$ is defined as the outer product $\mathbf{F}^{(t,u)} = \hat{\Delta^{-} \mathbf{b}_{u}^t} \otimes \Delta^{+} \mathbf{b}_{u}^t$, where $\hat{\Delta^{-} \mathbf{b}_{u}^t}$ is the $l_1$-normalized form of the absolute values in $\Delta^{-} \mathbf{b}_{u}^t$.
In this step, $f_{i,j}^{(t,u)}$ represents how much of the attention of user $u$ was transferred from subreddit $i$ to subreddit $j$ between the time-steps $t'$ and $t$.

\item By aggregating over users, we obtain the overall user's attention that migrated from subreddit $i$ to subreddit $j$ between the timesteps $t'$ and $t$:
\begin{equation*}
\mathbf{F}^{(t)} = \sum_{u\in \mathcal{U}} \mathbf{F}^{(t,u)}.
\end{equation*}

\item The AFG is defined as the aggregated network over multiple time-steps   $\mathcal{F} = \frac{1}{T}  \sum_{t=1}^{T} \mathbf{F}^{(t)}$.
\end{enumerate}

\begin{figure}
\centering
\includegraphics[width=1\linewidth]{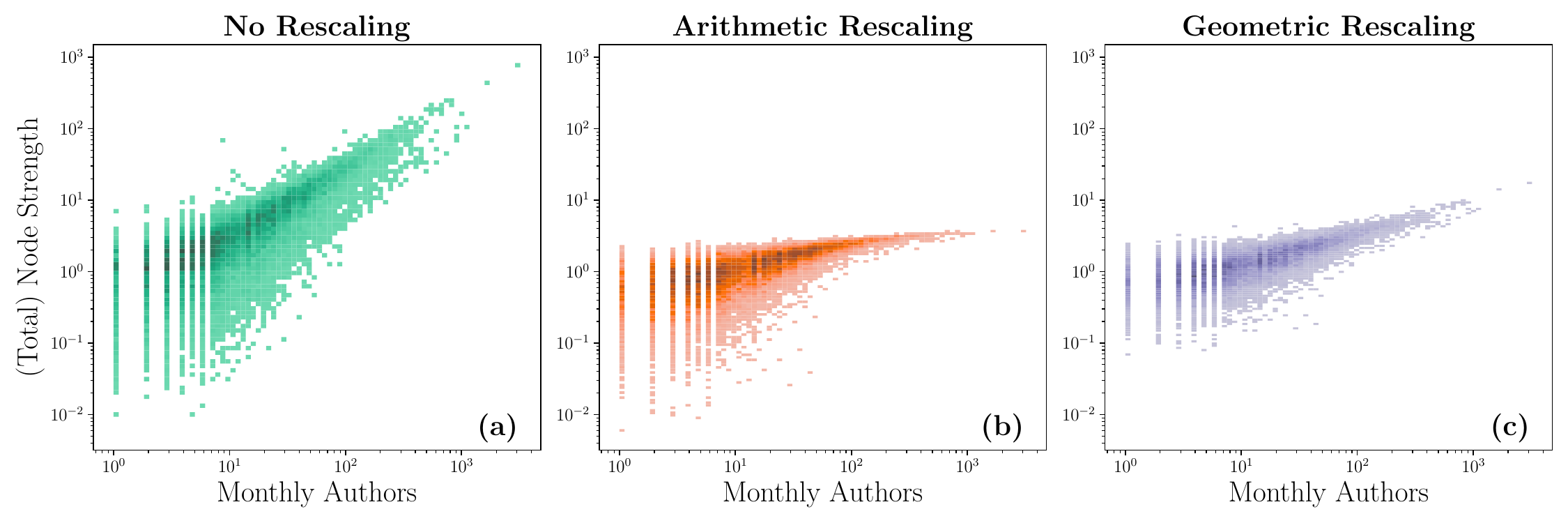}
\caption{\textbf{Weight Rescaling in the AFG.} Distribution of subreddits in January 2023 in the space defined by their number of monthly users and total node strength, \(s_i^{t} = s_{i}^{in}+s_{i}^{out}\). \textbf{(a)} shows the original distribution before weight rescaling, while \textbf{(b)} and \textbf{(c)} show the distributions obtained after rescaling edge weights by the arithmetic and geometric mean of the corresponding in/out node strengths, respectively. Both rescaling approaches reduce the correlation between node strength and subreddit popularity, although the geometric mean preserves the overall structure of the distribution more effectively while still mitigating this dependence.}
\label{fig:flow_rescaling}
\end{figure}

Following the original implementation, between steps 4 and 5, we rescale the network edge weights by the geometric mean of their in/out node strengths to dampen edge weight dependence on subreddits' popularity. \Cref{fig:flow_rescaling} 
illustrates the effect of weight rescaling on the dependence between 
node strength and subreddit popularity, motivating our choice of the 
geometric mean. 

\subsection{The Disparity Filter}\label{sm:disp_filter}

To remove non-significant edges from the AFG, we apply the Disparity Filter method proposed by Serrano et al. \cite{serranoExtractingMultiscaleBackbone2009}. This network backboning method identifies important edges by analyzing local weight heterogeneity around each node and assessing whether the observed heterogeneity can be explained by random fluctuations. Specifically, the method first normalizes edge weights as $p_{ij} = \omega_{ij}/s_i$, and then compares them against a null model in which weights are assigned uniformly at random. Under this hypothesis, the probability density of a normalized weight is:
\begin{equation*}
\rho(x) = (k-1)(1-x)^{k-2}.
\end{equation*}

Therefore, for each edge, a significance value is computed as:
\begin{equation*}
\alpha_{ij} = \mathds{P}(X\geq p_{ij}) = 1-\int_{0}^{p_{ij}} \rho(x)dx,
\end{equation*}

which represents the probability of observing a weight at least as large under the null model. Edges with $\alpha_{ij}<\alpha$ are considered statistically significant and retained, while the rest are discarded.

For directed networks, the procedure is applied separately to incoming and outgoing edges, producing $\alpha_{ij}^{in}$ and $\alpha_{ij}^{out}$. An edge is preserved in the backbone if it is significant for at least one of its two endpoint nodes.

\subsection{Statistics on the AFG}

\begin{figure}
  \centering
   \begin{subfigure}[b]{0.4\linewidth}
    \includegraphics[width=\linewidth]{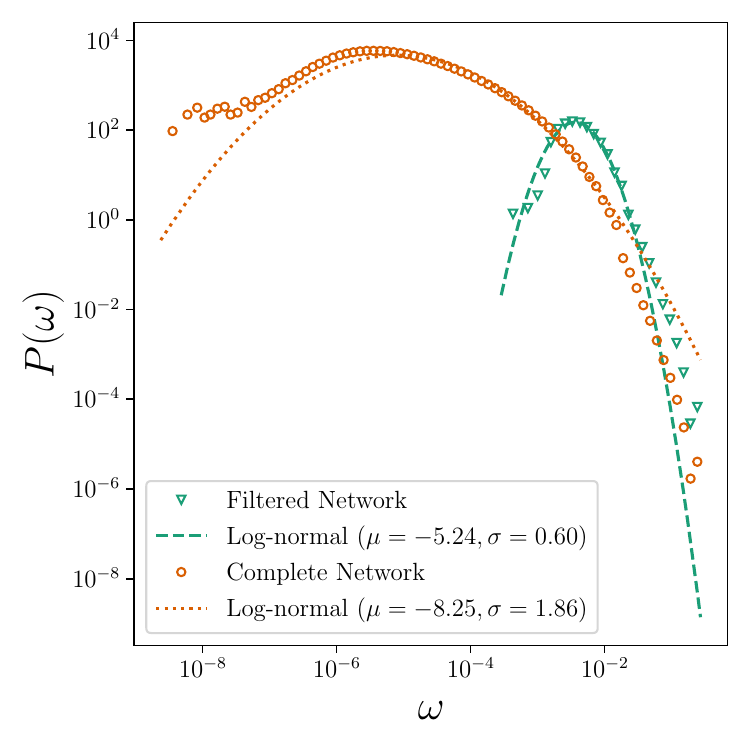}
     \phantomsubcaption\label{fig:weights_distribution}
     \vspace{-1.5\baselineskip} %
  \end{subfigure}
  \hspace{0.05\linewidth}
 \begin{subfigure}[b]{0.4\linewidth}
    \includegraphics[width=\linewidth]{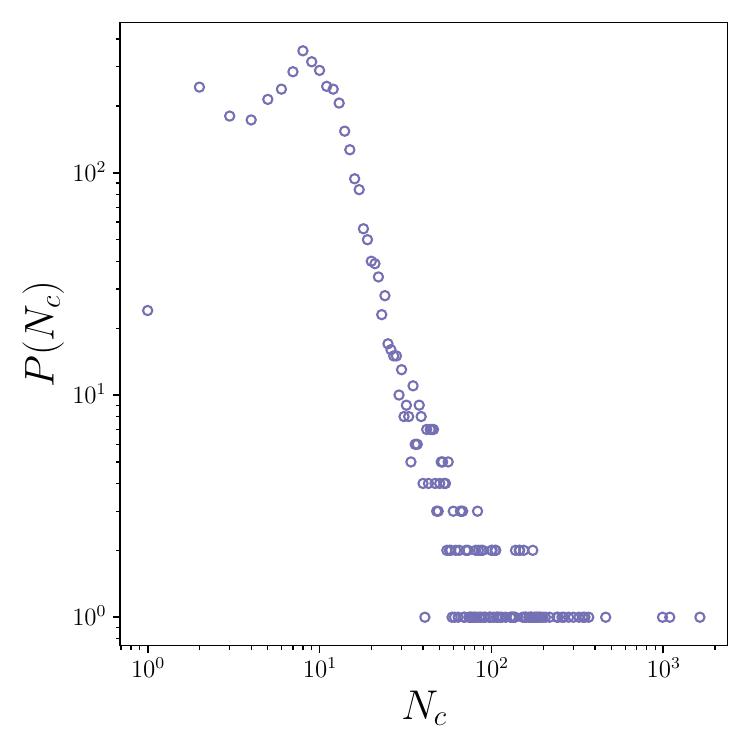}
    \phantomsubcaption\label{fig:community_size}
    \vspace{-1.5\baselineskip} %
  \end{subfigure}
  \caption{\textbf{AFG Weight and Community Size Distributions.} \textbf{(a)} shows the weight distribution before and after applying the Disparity Filter, with a likelihood-ratio test favoring a log-normal distribution over a power-law distribution in both cases. \textbf{(b)} shows the community size distribution of the communities found using the Infomap Algorithm. This distribution is power-law-like, with most communities having $\sim 10$ subreddits, and a few having higher community size.
  \label{fig:community-size}
  }
\end{figure}

\Cref{fig:weights_distribution} shows the edge weight 
distribution before and after sparsification with the Disparity 
Filter, while \Cref{fig:community_size} shows the size distribution 
of the communities detected by the Infomap algorithm. Finally, 
\Cref{tab:afg_summary} summarizes the five communities selected for 
the analysis, reporting their seed subreddits and the size of the 
corresponding induced subgraphs.

\begin{table}
\centering
\caption{\textbf{Summary statistics of the AFG communities.} Overview of the five communities analyzed in this study. For each community, we report the corresponding seed subreddit and the number of nodes $|\mathcal{N}|$ and intra-community edges $|\mathcal{L}|$ in the induced subgraph.}
\label{tab:afg_summary}
\large
\begin{tabular}{cccc}
\toprule
Community    & Seed & $|\mathcal{N}|$ & $|\mathcal{L}|$ \\
\midrule
Radical Left & \texttt{r/socialism} &133             & 817             \\ 
Esoterism    & \texttt{r/spirituality} &108             & 543             \\ 
Alt-Right    & \texttt{r/Conservative} &83              & 335             \\ 
Environment  & \texttt{r/environment} &43              & 115             \\ 
Conspiracy   & \texttt{r/conspiracy} &36              & 67              \\
\bottomrule
\end{tabular}
\end{table}

\section{Sociodemographic scores}
\label{si:sociodemographic-scores}
\begin{figure}
\centering
\includegraphics[width=\linewidth]{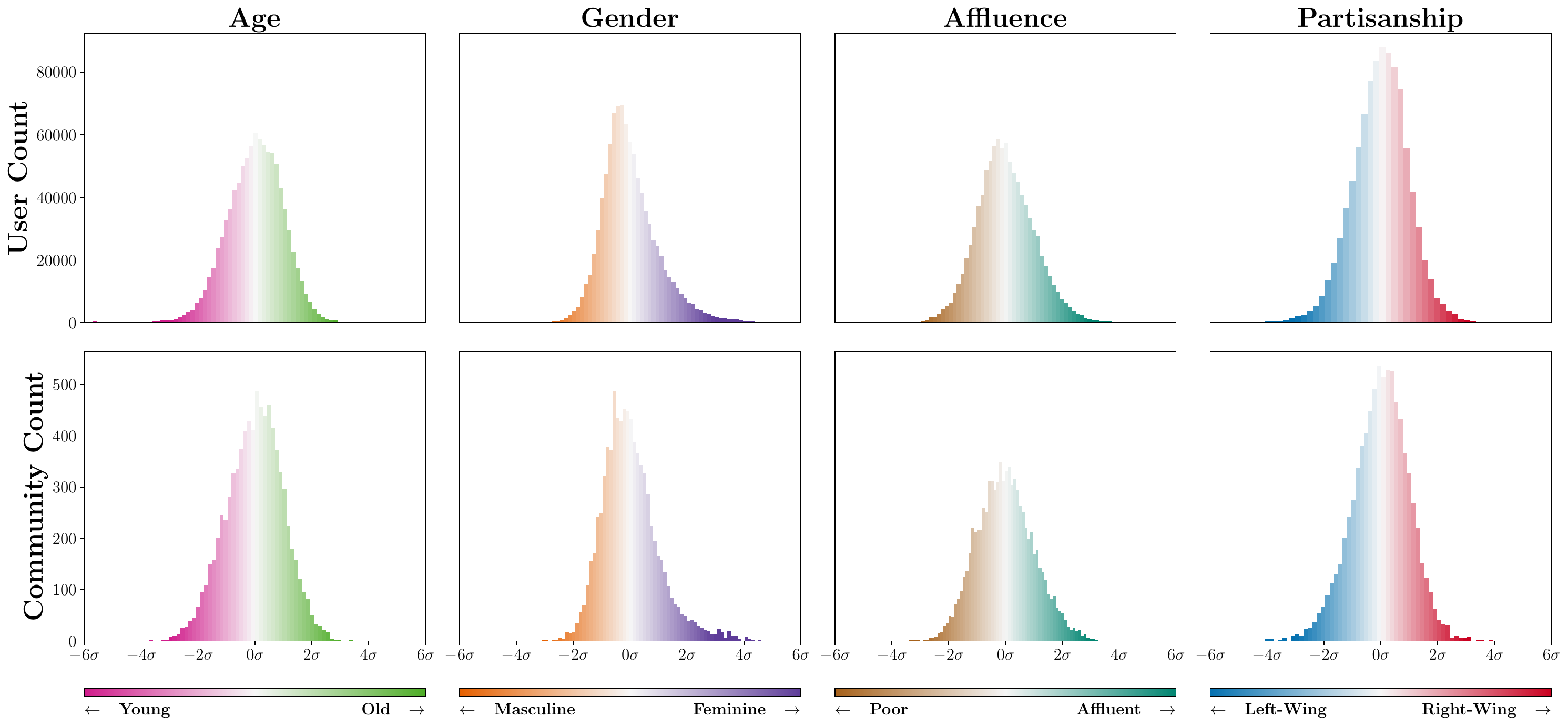}
\caption{\textbf{Score distributions across social dimensions.} Subreddit score distributions (bottom row) and the corresponding projected user score distributions (top row) on the age, gender, affluence, and partisanship social dimensions. The x-axis and the colors show the number of standard deviations from the mean score of each dimension (z-score).}
\label{fig:score-distributions}
\end{figure}

\begin{figure}
\centering
\includegraphics[width=0.45\linewidth]{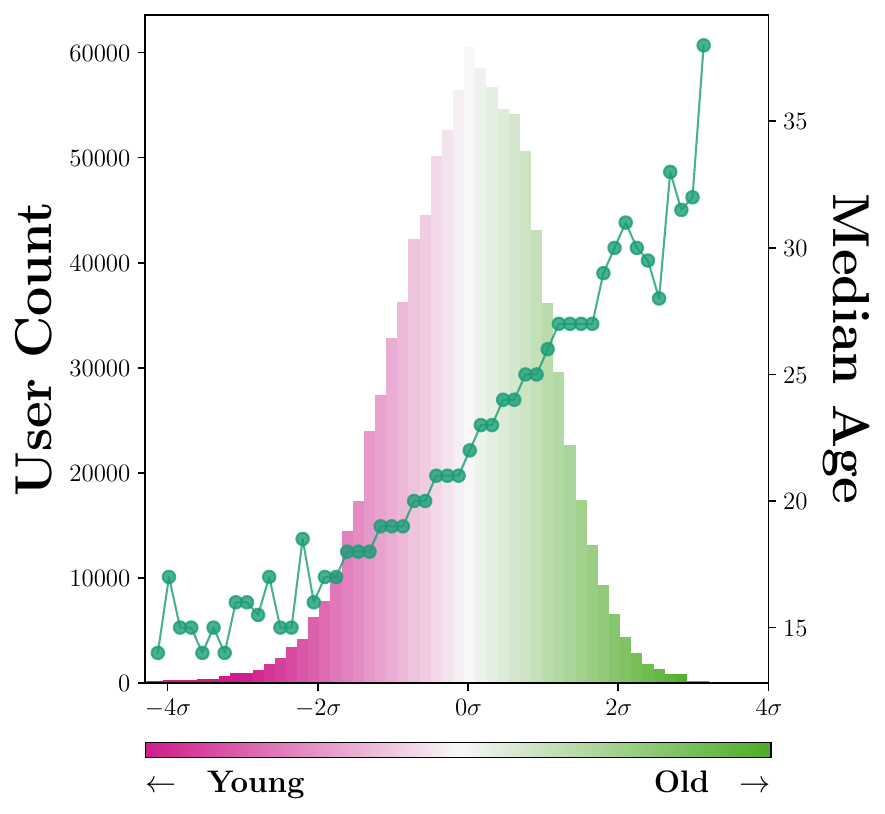}
\caption{\textbf{Projected vs self-declared age distribution.} Comparison between projected age scores and self-declared age within the same user cohort. We bin the self-declared age according to the z-score of the projected scores and report the median age for bins with at least 3 users. Self-reported age broadly follows a linear relationship as expected, with higher projected scores corresponding to older individuals, on average. Self-declared age data is shared by \citet{cinusUncoveringSociodemographicFabric2025}.}
\label{fig:age-comparison}
\end{figure}

\begin{figure}
\centering
\includegraphics[width=\linewidth]{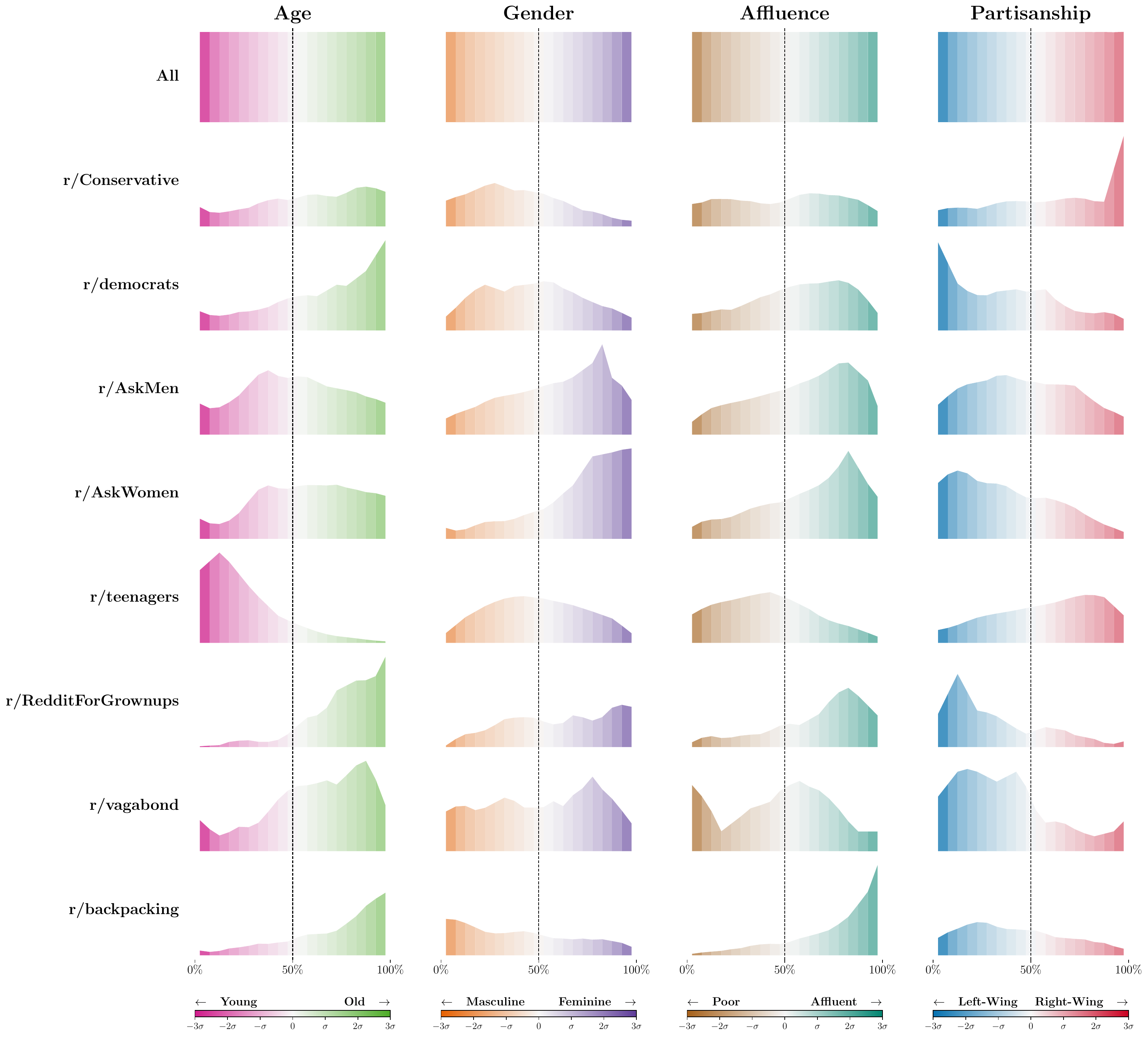}
\caption{\textbf{User score distributions across seed subreddits.}
User score distributions along age, gender, affluence, and partisanship dimensions prior to users joining the corresponding seed subreddits. The x-axis shows percentile-transformed user scores, and colour indicates z-scores of the distribution. The top row corresponds to all users and is, by construction, uniformly distributed on $U(0,100)$. Across seed subreddits, we assess the extent to which users are over- or under-represented along each social dimension. We observe clear separation along the age dimension between \texttt{r/teenagers} and \texttt{r/RedditForGrownups}, and along partisanship between \texttt{r/democrats} and \texttt{r/Conservative}. In contrast, gender and affluence show weaker or no clear separation, with \texttt{r/AskMen} and \texttt{r/vagabond} not exhibiting strong overrepresentation of the expected groups.}
\label{fig:quantile-density-reddit}
\end{figure}

Here, we report additional validation and descriptive statistics for the sociodemographic scores used throughout the analysis.
In \Cref{fig:score-distributions}, we show the distributions of subreddit- and user-level scores along the age, gender, affluence, and partisanship dimensions.
\Cref{fig:age-comparison} validates the projected age scores against self-declared age, showing that higher projected scores correspond, on average, to older users.
Finally, \Cref{fig:quantile-density-reddit} compares the pre-entry score distributions of users across the seed subreddits used to define each sociodemographic axis.

\section{Extended results on reachability, rankings, and URLs}
\label{si:extended-results}

\begin{figure}
\centering
\includegraphics[width=\linewidth]{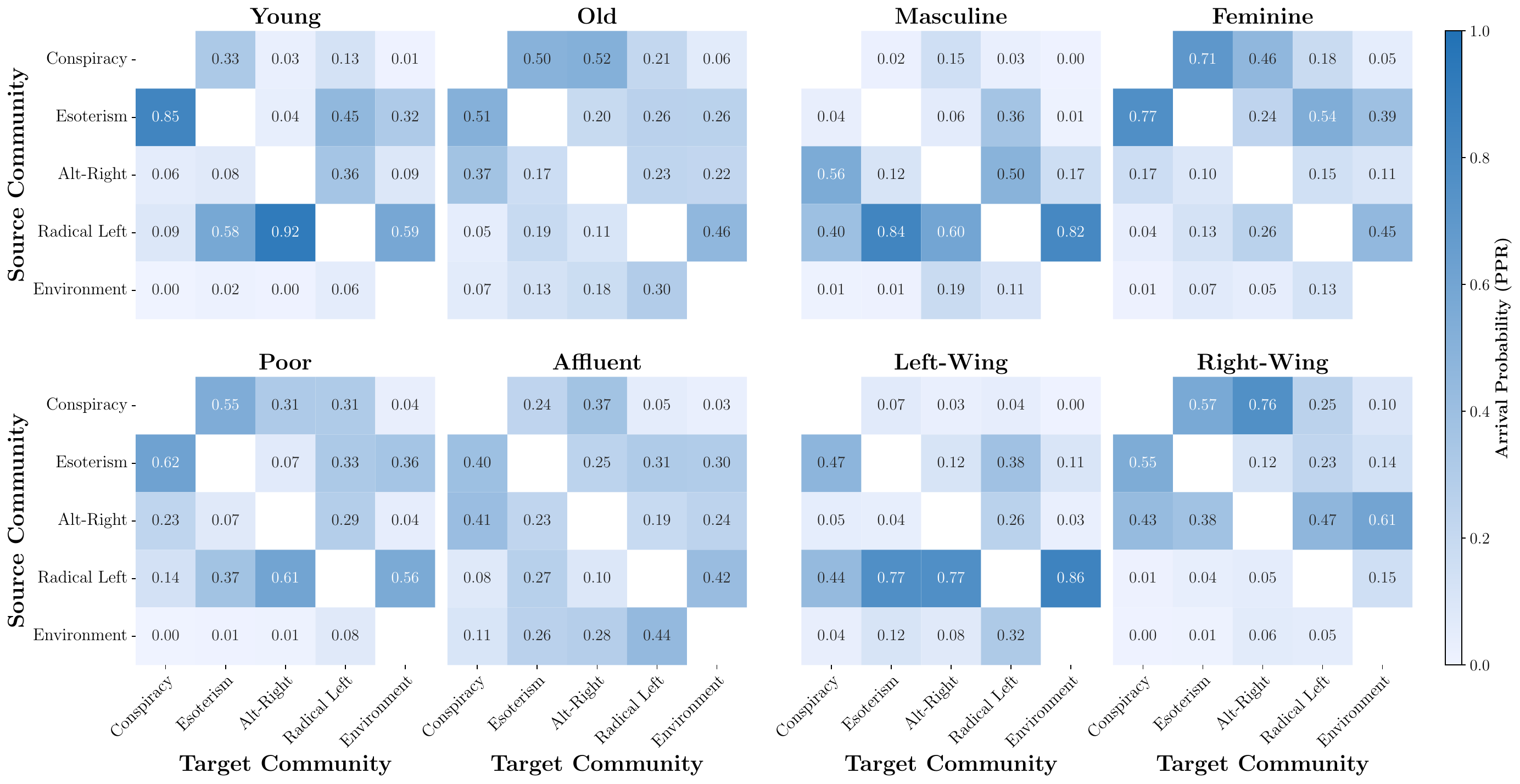}
\caption{\textbf{Cross-community reachability in stratified AFG.} Probability that a random-walk (with restart) reaches a target community $t$ if starting from a source community $s$ (with $t\neq s$) for each social dimension (age, gender, affluence, partisanship) and axis (positive/negative) of the stratified version of the AFG.}
\label{fig:reachability-stratified}
\end{figure}

\begin{figure}
    \centering
    \includegraphics[width=0.99\linewidth]{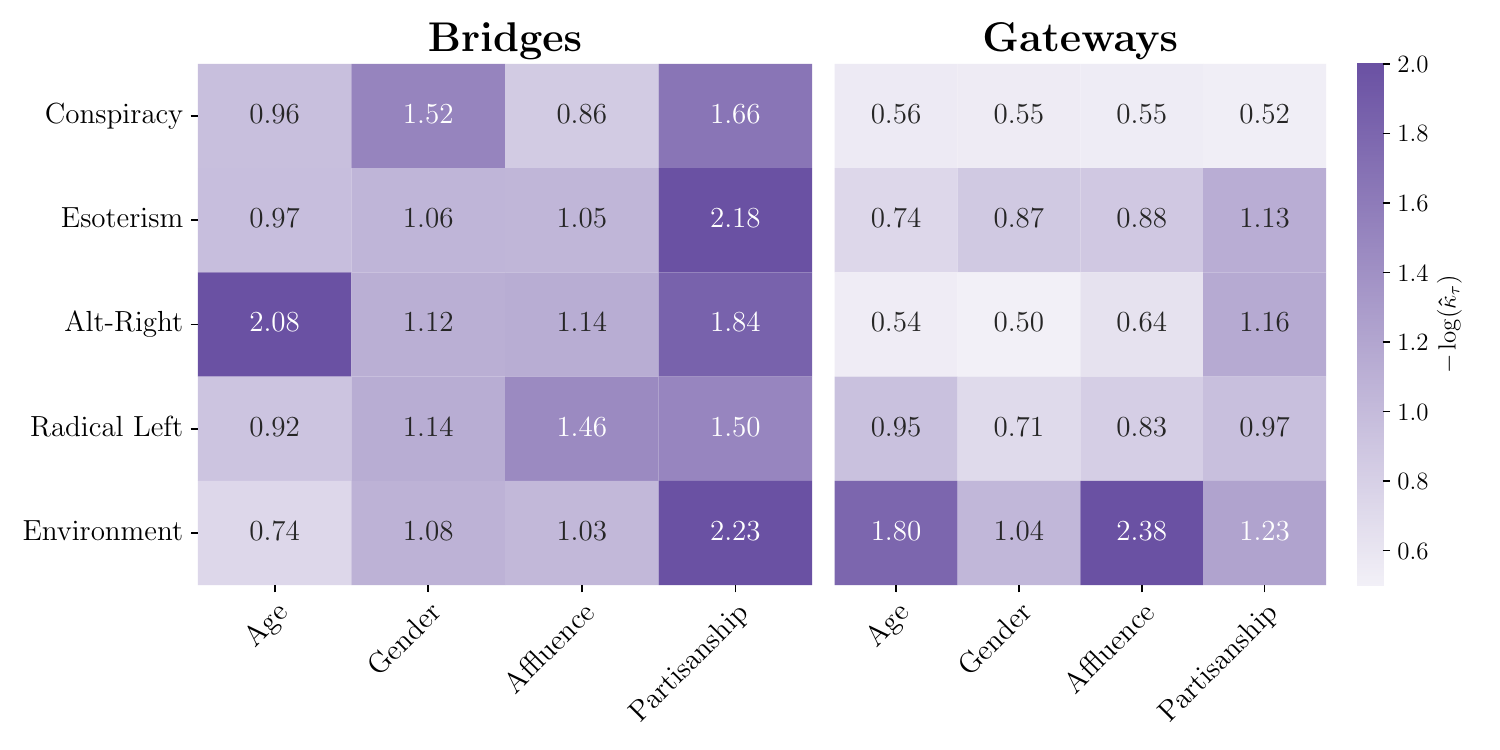}
    \caption{\textbf{Differences in bridge and gateway rankings across sociodemographic groups.} Negative logarithm of the weighted Kendall--Tau ($\hat{\kappa}_{\tau}$) values between the bridge and gateway rankings obtained from the positive and negative variants of the AFG for each social dimension. Higher values indicate greater differences in the rankings between sociodemographic groups.
    }
    \label{fig:kendall-tau-htmap}
\end{figure}

\begin{figure}
\centering
\makebox[\textwidth]{%
  \includegraphics[width=1.1\linewidth]{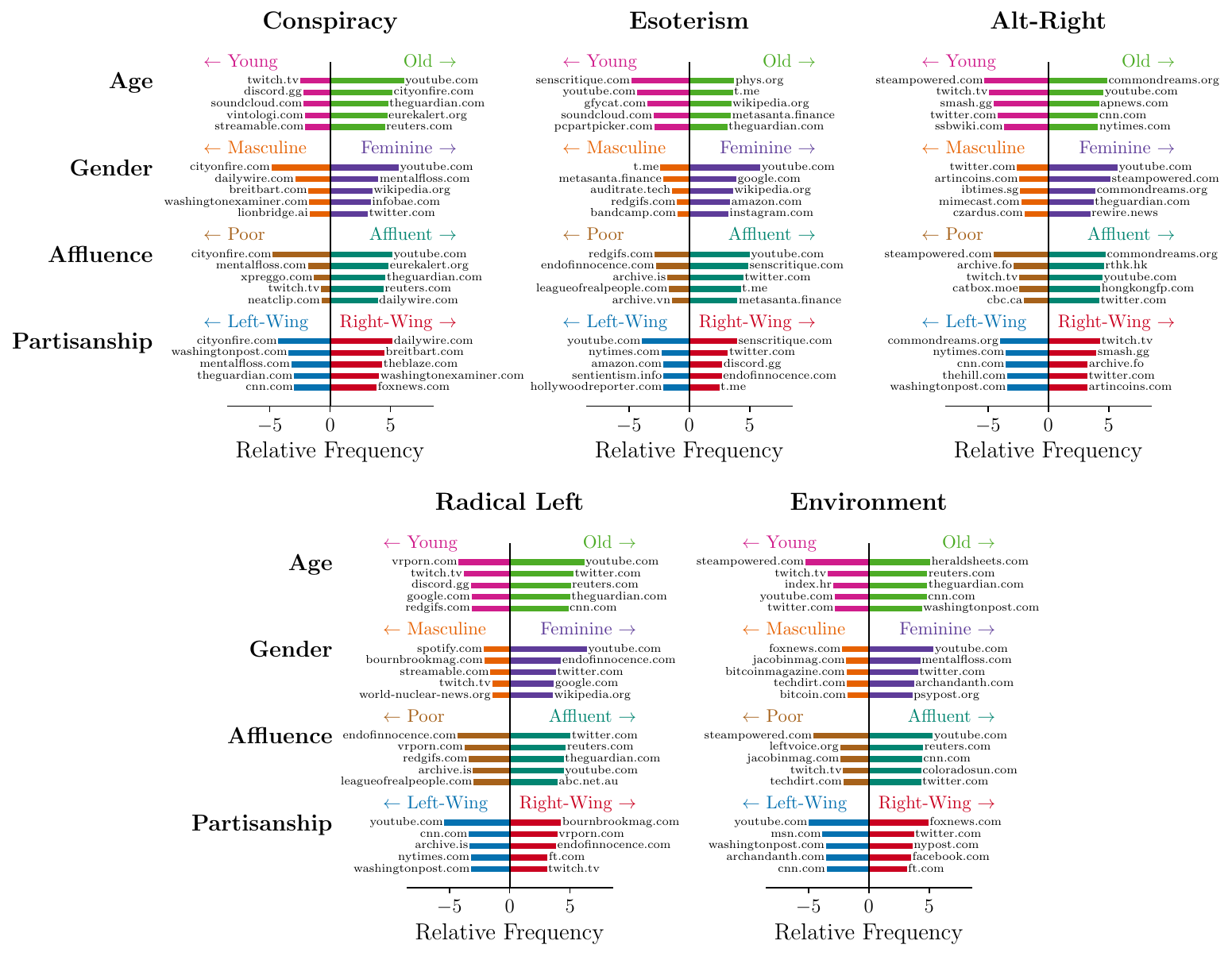
}}
\caption{\textbf{Most shared domains across sociodemographic groups.} Top five most shared URL domains by sociodemographic group prior to users entering political or conspiracy-related communities. For each social dimension $d$, we compute the ranking score $r_i^d$ of domain $i$ as a weighted sum of domain-sharing frequencies across users, weighted by the corresponding user scores in dimension $d$. A logarithmic transformation is subsequently applied to reduce the influence of outliers and highly skewed sharing distributions.}
\label{fig:url_sharing}
\end{figure}

Here, we provide the complete results for the reachability, bridge and gateway ranking, and URL-sharing analyses summarized in the main text.
\Cref{fig:reachability-stratified} reports cross-community reachability probabilities for both poles of each sociodemographic dimension.
\Cref{fig:kendall-tau-htmap} compares gateway and bridge rankings between these poles, with larger values indicating greater differences between groups. 
In \Cref{fig:url_sharing}, we show the external domains most frequently shared by users from each sociodemographic group before they enter the communities under study.

\section{Bridges and gateways full rankings}

Here, we report the full gateway and bridge rankings underlying the aggregate comparisons presented in the main text. 
\Cref{tab:conspiracy_gateways,tab:conservative_gateways,tab:environment_gateways,tab:esoterism_gateways,tab:radical_left_gateways} list the highest-ranked gateways within each community for both poles of every sociodemographic dimension.
\Cref{tab:conspiracy_bridges,tab:conservative_bridges,tab:environment_bridges,tab:radical_left_bridges,tab:esoterism_bridges} report the corresponding bridge rankings, identifying the subreddits that lie on incoming pathways toward each community.

\begingroup
\setlength{\tabcolsep}{5pt}

\begin{table}
\centering
\caption{\textbf{Top gateway subreddits to conspiracy content across sociodemographic axes.}
Top three gateway subreddits ranked by personalized PageRank (PPR) for the Conspiracy community across each sociodemographic dimension and axis.}
\label{tab:conspiracy_gateways}
\small
\begin{adjustbox}{max width=\textwidth}
\begin{tabular}{ccc}
\toprule
\multirow{2}{*}{\textbf{Social Dimension}} & \multicolumn{2}{c}{\textbf{Top-3 Gateways}}               \\ \cmidrule(lr){2-3} 
& \multicolumn{1}{c}{\textbf{Negative}}                  & \textbf{Positive}\\ \midrule
\multirow{3}{*}{\textbf{\begin{tabular}[c]{@{}c@{}}Age\\ (Young - Old)\end{tabular}}} &
  \multicolumn{1}{c}{\texttt{r/conspiracytheories}} &
  \texttt{r/conspiracy} \\& \multicolumn{1}{c}{\texttt{r/ConspiracyMemes}}     & \texttt{r/conspiracy\_commons} \\& \multicolumn{1}{c}{\texttt{r/conspiracy}}& \texttt{r/conspiracytheories}  \\ \midrule
\multirow{3}{*}{\textbf{\begin{tabular}[c]{@{}c@{}}Gender\\ (Masculine - Feminine)\end{tabular}}} &
  \multicolumn{1}{c}{\texttt{r/conspiracy\_commons}} &
  \texttt{r/conspiracy} \\& \multicolumn{1}{c}{\texttt{r/conspiracytheories}}  & \texttt{r/conspiracytheories}  \\ & \multicolumn{1}{c}{\texttt{r/conspiracy}}          & \texttt{r/conspiracy\_commons} \\ \midrule
\multirow{3}{*}{\textbf{\begin{tabular}[c]{@{}c@{}}Affluence\\ (Poor - Affluent)\end{tabular}}} &
  \multicolumn{1}{c}{\texttt{r/conspiracy}} &
  \texttt{r/conspiracy} \\& \multicolumn{1}{c}{\texttt{r/conspiracy\_commons}} & \texttt{r/conspiracytheories}  \\ & \multicolumn{1}{c}{\texttt{r/ConspiracyMemes}}     & \texttt{r/conspiracy\_commons} \\ \midrule
\multirow{3}{*}{\textbf{\begin{tabular}[c]{@{}c@{}}Partisan\\ (Left-Wing - Right-Wing)\end{tabular}}} &
  \multicolumn{1}{c}{\texttt{r/EscapingPrisonPlanet}} &
  \texttt{r/conspiracy} \\ & \multicolumn{1}{c}{\texttt{r/conspiracy\_commons}} & \texttt{r/conspiracy\_commons} \\ & \multicolumn{1}{c}{\texttt{r/conspiracy}}         & \texttt{r/conspiracytheories}  \\ \bottomrule
\end{tabular}
\end{adjustbox}
\end{table}

\begin{table}
\centering
\caption{\textbf{Top gateway subreddits to alt-right content across sociodemographic axes.}
Top three gateway subreddits ranked by personalized PageRank (PPR) for the Alt-Right community across each sociodemographic dimension and axis.}
\label{tab:conservative_gateways}
\small
\begin{adjustbox}{max width=\textwidth}
\begin{tabular}{ccc}
\toprule
\multirow{2}{*}{\textbf{Social Dimension}} & \multicolumn{2}{c}{\textbf{Top-3 Gateways}}               \\ \cmidrule(lr){2-3} 
& \multicolumn{1}{c}{\textbf{Negative}}                  & \textbf{Positive}\\ \midrule
\multirow{3}{*}{\textbf{\begin{tabular}[c]{@{}c@{}}Age\\ (Young - Old)\end{tabular}}} &
  \multicolumn{1}{c}{\texttt{r/The\_Donald}} &
  \texttt{r/Conservative} \\& \multicolumn{1}{c}{\texttt{r/LouderWithCrowder}}     & \texttt{r/JordanPeterson} \\& \multicolumn{1}{c}{\texttt{r/JordanPeterson}}& \texttt{r/WayOfTheBern}  \\ \midrule
\multirow{3}{*}{\textbf{\begin{tabular}[c]{@{}c@{}}Gender\\ (Masculine - Feminine)\end{tabular}}} &
  \multicolumn{1}{c}{\texttt{r/JordanPeterson}} &
  \texttt{r/CoronavirusCirclejerk} \\& \multicolumn{1}{c}{\texttt{r/Conservative}}  & \texttt{r/Conservative}  \\ & \multicolumn{1}{c}{\texttt{r/The\_Donald}}          & \texttt{r/WayOfTheBern} \\ \midrule
\multirow{3}{*}{\textbf{\begin{tabular}[c]{@{}c@{}}Affluence\\ (Poor - Affluent)\end{tabular}}} &
  \multicolumn{1}{c}{\texttt{r/The\_Donald}} &
  \texttt{r/Conservative} \\& \multicolumn{1}{c}{\texttt{r/WayOfTheBern}} & \texttt{r/JordanPeterson}  \\ & \multicolumn{1}{c}{\texttt{r/ShitPoliticsSays}}     & \texttt{r/Republican} \\ \midrule
\multirow{3}{*}{\textbf{\begin{tabular}[c]{@{}c@{}}Partisan\\ (Left-Wing - Right-Wing)\end{tabular}}} &
  \multicolumn{1}{c}{\texttt{r/WayOfTheBern}} &
  \texttt{r/Conservative} \\ & \multicolumn{1}{c}{\texttt{r/Trumpgret}} & \texttt{r/The\_Donald} \\ & \multicolumn{1}{c}{\texttt{r/LockdownSkepticism}}         & \texttt{r/JordanPeterson}  \\ \bottomrule
\end{tabular}
\end{adjustbox}
\end{table}
\begin{table}
\centering
\caption{\textbf{Top gateway subreddits to environmental content across sociodemographic axes.}
Top three gateway subreddits ranked by personalized PageRank (PPR) for the Environment community across each sociodemographic dimension and axis.}
\label{tab:environment_gateways}
\small
\begin{adjustbox}{max width=\textwidth}
\begin{tabular}{ccc}
\toprule
\multirow{2}{*}{ 	\textbf{Social Dimension}} & \multicolumn{2}{c}{ 	\textbf{Top-3 Gateways}}               \\ \cmidrule(lr){2-3} 
& \multicolumn{1}{c}{ 	\textbf{Negative}}                  &  	\textbf{Positive}\\ \midrule
\multirow{3}{*}{\textbf{\begin{tabular}[c]{@{}c@{}} 	Age\\  	(Young - Old)\end{tabular}}} &
  \multicolumn{1}{c}{	\texttt{r/climatechange}} &
   	\texttt{r/environment} \\& \multicolumn{1}{c}{ 	 \texttt{r/ClimateOffensive}}     &  	\texttt{r/energy} \\& \multicolumn{1}{c}{ 	\texttt{r/economy}}&  	\texttt{r/business}  \\ \midrule
\multirow{3}{*}{\textbf{\begin{tabular}[c]{@{}c@{}} 	Gender\\  	(Masculine - Feminine)\end{tabular}}} &
  \multicolumn{1}{c}{ 	\texttt{r/Economics}} &
   	\texttt{r/ZeroWaste} \\& \multicolumn{1}{c}{ 	\texttt{r/economy}}  &  	\texttt{r/environment}  \\ & \multicolumn{1}{c}{ 	\texttt{r/ClimateActionPlan}}          &  	\texttt{r/climate} \\ \midrule
\multirow{3}{*}{\textbf{\begin{tabular}[c]{@{}c@{}} 	Affluence\\  	(Poor - Affluent)\end{tabular}}} &
  \multicolumn{1}{c}{ 	\texttt{r/climate}} &
   	\texttt{r/environment} \\& \multicolumn{1}{c}{ 	\texttt{r/climatechange}} &  	\texttt{r/energy}  \\ & \multicolumn{1}{c}{ 	\texttt{r/ClimateOffensive}}     &  	\texttt{r/ZeroWaste} \\ \midrule
\multirow{3}{*}{\textbf{\begin{tabular}[c]{@{}c@{}} 	Partisan\\  	(Left-Wing - Right-Wing)\end{tabular}}} &
  \multicolumn{1}{c}{ 	\texttt{r/environment}} &
   	\texttt{r/economy} \\ & \multicolumn{1}{c}{ 	\texttt{r/ZeroWaste}} &  	\texttt{r/Economics} \\ & \multicolumn{1}{c}{\texttt{r/climate}}         & \texttt{r/business}  \\ \bottomrule
\end{tabular}
\end{adjustbox}
\end{table}
\begin{table}
\centering
\caption{\textbf{Top gateway subreddits to esoteric content across sociodemographic axes.}
Top three gateway subreddits ranked by personalized PageRank (PPR) for the Esoterism community across each sociodemographic dimension and axis.}
\label{tab:esoterism_gateways}
\small
\begin{adjustbox}{max width=\textwidth}
\begin{tabular}{ccc}
\toprule
\multirow{2}{*}{\textbf{Social Dimension}} & \multicolumn{2}{c}{\textbf{Top-3 Gateways}}               \\ \cmidrule(lr){2-3} 
& \multicolumn{1}{c}{\textbf{Negative}}                  & \textbf{Positive}\\ \midrule
\multirow{3}{*}{\textbf{\begin{tabular}[c]{@{}c@{}}Age\\ (Young - Old)\end{tabular}}} &
  \multicolumn{1}{c}{\texttt{r/Dreams}} &
  \texttt{r/spirituality} \\& \multicolumn{1}{c}{\texttt{r/LucidDreaming}}     & \texttt{r/Meditation} \\& \multicolumn{1}{c}{\texttt{r/AstralProjection}}& \texttt{r/tarot}  \\ \midrule
\multirow{3}{*}{\textbf{\begin{tabular}[c]{@{}c@{}}Gender\\ (Masculine - Feminine)\end{tabular}}} &
  \multicolumn{1}{c}{\texttt{r/Dreams}} &
  \texttt{r/Dreams} \\& \multicolumn{1}{c}{\texttt{r/GetMotivated}}  & \texttt{r/witchcraft}  \\ & \multicolumn{1}{c}{\texttt{r/confidence}}          & \texttt{r/tarot} \\ \midrule
\multirow{3}{*}{\textbf{\begin{tabular}[c]{@{}c@{}}Affluence\\ (Poor - Affluent)\end{tabular}}} &
  \multicolumn{1}{c}{\texttt{r/AstralProjection}} &
  \texttt{r/Meditation} \\& \multicolumn{1}{c}{\texttt{r/Retconned}} & \texttt{r/spirituality}  \\ & \multicolumn{1}{c}{\texttt{r/occult}}     & \texttt{r/WitchesVsPatriarchy} \\ \midrule
\multirow{3}{*}{\textbf{\begin{tabular}[c]{@{}c@{}}Partisan\\ (Left-Wing - Right-Wing)\end{tabular}}} &
  \multicolumn{1}{c}{\texttt{r/tarot}} &
  \texttt{r/LucidDreaming} \\ & \multicolumn{1}{c}{\texttt{r/witchcraft}} & \texttt{r/AstralProjection} \\ & \multicolumn{1}{c}{\texttt{r/WitchesVsPatriarchy}}         & \texttt{r/Paranormal}  \\ \bottomrule
\end{tabular}
\end{adjustbox}
\end{table}
\begin{table}
\centering
\caption{\textbf{Top gateway subreddits to radical left content across sociodemographic axes.}
Top three gateway subreddits ranked by personalized PageRank (PPR) for the Radical Left community across each sociodemographic dimension and axis.}
\label{tab:radical_left_gateways}
\small
\begin{adjustbox}{max width=\textwidth}
\begin{tabular}{ccc}
\toprule
\multirow{2}{*}{\textbf{Social Dimension}} & \multicolumn{2}{c}{\textbf{Top-3 Gateways}}               \\ \cmidrule(lr){2-3} 
& \multicolumn{1}{c}{\textbf{Negative}}                  & \textbf{Positive}\\ \midrule
\multirow{3}{*}{\textbf{\begin{tabular}[c]{@{}c@{}}Age\\ (Young - Old)\end{tabular}}} &
  \multicolumn{1}{c}{\texttt{r/communism}} &
  \texttt{r/LateStageCapitalism} \\& \multicolumn{1}{c}{\texttt{r/communism101}}     & \texttt{r/lostgeneration} \\& \multicolumn{1}{c}{\texttt{r/ShitLiberalsSay}}& \texttt{r/PropagandaPosters}  \\ \midrule
\multirow{3}{*}{\textbf{\begin{tabular}[c]{@{}c@{}}Gender\\ (Masculine - Feminine)\end{tabular}}} &
  \multicolumn{1}{c}{\texttt{r/ChapoTrapHouse}} &
  \texttt{r/LateStageCapitalism} \\& \multicolumn{1}{c}{\texttt{r/chapotraphouse2}}  & \texttt{r/communism}  \\ & \multicolumn{1}{c}{\texttt{r/leftistvexillology}}          & \texttt{r/Anarchy101} \\ \midrule
\multirow{3}{*}{\textbf{\begin{tabular}[c]{@{}c@{}}Affluence\\ (Poor - Affluent)\end{tabular}}} &
  \multicolumn{1}{c}{\texttt{r/communism}} &
  \texttt{r/LateStageCapitalism} \\& \multicolumn{1}{c}{\texttt{r/Anarchy101}} & \texttt{r/Sino}  \\ & \multicolumn{1}{c}{\texttt{r/communism101}}     & \texttt{r/GreenAndPleasant} \\ \midrule
\multirow{3}{*}{\textbf{\begin{tabular}[c]{@{}c@{}}Partisan\\ (Left-Wing - Right-Wing)\end{tabular}}} &
  \multicolumn{1}{c}{\texttt{r/ChapoTrapHouse}} &
  \texttt{r/DebateCommunism} \\ & \multicolumn{1}{c}{\texttt{r/communism}} & \texttt{r/CommunismMemes} \\ & \multicolumn{1}{c}{\texttt{r/communism101}}         & \texttt{r/LateStageCapitalism}  \\ \bottomrule
\end{tabular}
\end{adjustbox}
\end{table}
\begin{table}
\centering
\caption{\textbf{Top bridge subreddits to conspiracy content across sociodemographic axes.}
Top three bridge subreddits ranked by personalized PageRank (PPR) for the Conspiracy community across each sociodemographic dimension and axis. We report each subreddit's community membership below their name.}
\label{tab:conspiracy_bridges}
\small
\begin{adjustbox}{max width=\textwidth}
\begin{tabular}{ccc}
\toprule
\multirow{2}{*}{\textbf{Social Dimension}} &
  \multicolumn{2}{c}{\textbf{Top-3 Bridges}} \\ \cmidrule(lr){2-3}
 &
  \multicolumn{1}{c}{\textbf{Negative}} &
  \textbf{Positive} \\ \midrule
\multirow{3}{*}{\textbf{\begin{tabular}[c]{@{}c@{}}Age\\ (Young - Old)\end{tabular}}} &
  \multicolumn{1}{c}{\begin{tabular}[c]{@{}c@{}}\texttt{r/AstralProjection}\\ {\footnotesize (Esoterism)}\end{tabular}} &
  \begin{tabular}[c]{@{}c@{}}\texttt{r/Retconned}\\ {\footnotesize (Esoterism)}\end{tabular} \\ 
 &
  \multicolumn{1}{c}{\begin{tabular}[c]{@{}c@{}}\texttt{r/LucidDreaming}\\ {\footnotesize (Esoterism)}\end{tabular}} &
  \begin{tabular}[c]{@{}c@{}}\texttt{r/awakened}\\ {\footnotesize (Esoterism)}\end{tabular} \\ 
 &
  \multicolumn{1}{c}{\begin{tabular}[c]{@{}c@{}}\texttt{r/Wicca}\\ {\footnotesize (Esoterism)}\end{tabular}} &
  \begin{tabular}[c]{@{}c@{}}\texttt{r/NoNewNormal}\\ {\footnotesize (Alt-Right)}\end{tabular} \\ \midrule
\multirow{3}{*}{\textbf{\begin{tabular}[c]{@{}c@{}}Gender\\ (Masculine - Feminine)\end{tabular}}} &
  \multicolumn{1}{c}{\begin{tabular}[c]{@{}c@{}}\texttt{r/CoronavirusCirclejerk}\\ {\footnotesize (Alt-Right)}\end{tabular}} &
  \begin{tabular}[c]{@{}c@{}}\texttt{r/Retconned}\\ {\footnotesize (Esoterism)}\end{tabular} \\ 
 &
  \multicolumn{1}{c}{\begin{tabular}[c]{@{}c@{}}\texttt{r/Paranormal}\\ {\footnotesize (Esoterism)}\end{tabular}} &
  \begin{tabular}[c]{@{}c@{}}\texttt{r/Wicca}\\ {\footnotesize (Esoterism)}\end{tabular} \\ 
 &
  \multicolumn{1}{c}{\begin{tabular}[c]{@{}c@{}}\texttt{r/ChapoTrapHouse}\\ {\footnotesize (Radical Left)}\end{tabular}} &
  \begin{tabular}[c]{@{}c@{}}\texttt{r/AstralProjection}\\ {\footnotesize (Esoterism)}\end{tabular} \\ \midrule
\multirow{3}{*}{\textbf{\begin{tabular}[c]{@{}c@{}}Affluence\\ (Poor - Affluent)\end{tabular}}} &
  \multicolumn{1}{c}{\begin{tabular}[c]{@{}c@{}}\texttt{r/Retconned}\\ {\footnotesize (Esoterism)}\end{tabular}} &
  \begin{tabular}[c]{@{}c@{}}\texttt{r/CoronavirusCirclejerk}\\ {\footnotesize (Alt-Right)}\end{tabular} \\ 
 &
  \multicolumn{1}{c}{\begin{tabular}[c]{@{}c@{}}\texttt{r/AstralProjection}\\ {\footnotesize (Esoterism)}\end{tabular}} &
  \begin{tabular}[c]{@{}c@{}}\texttt{r/Meditation}\\ {\footnotesize (Esoterism)}\end{tabular} \\ 
 &
  \multicolumn{1}{c}{\begin{tabular}[c]{@{}c@{}}\texttt{r/occult}\\ {\footnotesize (Esoterism)}\end{tabular}} &
  \begin{tabular}[c]{@{}c@{}}\texttt{r/FreeSpeech}\\ {\footnotesize (Alt-Right)}\end{tabular} \\ \midrule
\multirow{3}{*}{\textbf{\begin{tabular}[c]{@{}c@{}}Partisan\\ (Left-Wing - Right-Wing)\end{tabular}}} &
  \multicolumn{1}{c}{\begin{tabular}[c]{@{}c@{}}\texttt{r/Wicca}\\ {\footnotesize (Esoterism)}\end{tabular}} &
  \begin{tabular}[c]{@{}c@{}}\texttt{r/AstralProjection}\\ {\footnotesize (Esoterism)}\end{tabular} \\ 
 &
  \multicolumn{1}{c}{\begin{tabular}[c]{@{}c@{}}\texttt{r/tarot}\\ {\footnotesize (Esoterism)}\end{tabular}} &
  \begin{tabular}[c]{@{}c@{}}\texttt{r/LucidDreaming}\\ {\footnotesize (Esoterism)}\end{tabular} \\ 
 &
  \multicolumn{1}{c}{\begin{tabular}[c]{@{}c@{}}\texttt{r/witchcraft}\\ {\footnotesize (Esoterism)}\end{tabular}} &
  \begin{tabular}[c]{@{}c@{}}\texttt{r/Retconned}\\ {\footnotesize (Esoterism)}\end{tabular} \\ \bottomrule
\end{tabular}
\end{adjustbox}
\end{table}
\begin{table}
\centering
\caption{\textbf{Top bridge subreddits to alt-right content across sociodemographic axes.}
Top three bridge subreddits ranked by personalized PageRank (PPR) for the Alt-Right community across each sociodemographic dimension and axis. We report each subreddit's community membership below their name.}
\label{tab:conservative_bridges}
\small
\begin{adjustbox}{max width=\textwidth}
\begin{tabular}{ccc}
\toprule
\multirow{2}{*}{\textbf{Social Dimension}} &
  \multicolumn{2}{c}{\textbf{Top-3 Bridges}} \\ \cmidrule(lr){2-3}
 &
  \multicolumn{1}{c}{\textbf{Negative}} &
  \textbf{Positive} \\ \midrule
\multirow{3}{*}{\textbf{\begin{tabular}[c]{@{}c@{}}Age\\ (Young - Old)\end{tabular}}} &
  \multicolumn{1}{c}{\begin{tabular}[c]{@{}c@{}}\texttt{r/ShitLiberalsSay}\\ {\footnotesize (Radical Left)}\end{tabular}} &
  \begin{tabular}[c]{@{}c@{}}\texttt{r/conspiracy}\\ {\footnotesize (Conspiracy)}\end{tabular} \\ 
 &
  \multicolumn{1}{c}{\begin{tabular}[c]{@{}c@{}}\texttt{r/GenZedong}\\ {\footnotesize (Radical Left)}\end{tabular}} &
  \begin{tabular}[c]{@{}c@{}}\texttt{r/ConspiracyMemes}\\ {\footnotesize (Conspiracy)}\end{tabular} \\ 
 &
  \multicolumn{1}{c}{\begin{tabular}[c]{@{}c@{}}\texttt{r/Anarchy101}\\ {\footnotesize (Radical Left)}\end{tabular}} &
  \begin{tabular}[c]{@{}c@{}}\texttt{r/energy}\\ {\footnotesize (Environment)}\end{tabular} \\ \midrule
\multirow{3}{*}{\textbf{\begin{tabular}[c]{@{}c@{}}Gender\\ (Masculine - Feminine)\end{tabular}}} &
  \multicolumn{1}{c}{\begin{tabular}[c]{@{}c@{}}\texttt{r/economy}\\ {\footnotesize (Environment)}\end{tabular}} &
  \begin{tabular}[c]{@{}c@{}}\texttt{r/conspiracy}\\ {\footnotesize (Conspiracy)}\end{tabular} \\ 
 &
  \multicolumn{1}{c}{\begin{tabular}[c]{@{}c@{}}\texttt{r/ChapoTrapHouse}\\ {\footnotesize (Radical Left)}\end{tabular}} &
  \begin{tabular}[c]{@{}c@{}}\texttt{r/ShitLiberalsSay}\\ {\footnotesize (Radical Left)}\end{tabular} \\ 
 &
  \multicolumn{1}{c}{\begin{tabular}[c]{@{}c@{}}\texttt{r/ShitLiberalsSay}\\ {\footnotesize (Radical Left)}\end{tabular}} &
  \begin{tabular}[c]{@{}c@{}}\texttt{r/ConspiracyMemes}\\ {\footnotesize (Conspiracy)}\end{tabular} \\ \midrule
\multirow{3}{*}{\textbf{\begin{tabular}[c]{@{}c@{}}Affluence\\ (Poor - Affluent)\end{tabular}}} &
  \multicolumn{1}{c}{\begin{tabular}[c]{@{}c@{}}\texttt{r/conspiracy}\\ {\footnotesize (Conspiracy)}\end{tabular}} &
  \begin{tabular}[c]{@{}c@{}}\texttt{r/conspiracy}\\ {\footnotesize (Conspiracy)}\end{tabular} \\ 
 &
  \multicolumn{1}{c}{\begin{tabular}[c]{@{}c@{}}\texttt{r/ShitLiberalsSay}\\ {\footnotesize (Radical Left)}\end{tabular}} &
  \begin{tabular}[c]{@{}c@{}}\texttt{r/energy}\\ {\footnotesize (Environment)}\end{tabular} \\ 
 &
  \multicolumn{1}{c}{\begin{tabular}[c]{@{}c@{}}\texttt{r/ConspiracyMemes}\\ {\footnotesize (Conspiracy)}\end{tabular}} &
  \begin{tabular}[c]{@{}c@{}}\texttt{r/economy}\\ {\footnotesize (Environment)}\end{tabular} \\ \midrule
\multirow{3}{*}{\textbf{\begin{tabular}[c]{@{}c@{}}Partisan\\ (Left-Wing - Right-Wing)\end{tabular}}} &
  \multicolumn{1}{c}{\begin{tabular}[c]{@{}c@{}}\texttt{r/ShitLiberalsSay}\\ {\footnotesize (Radical Left)}\end{tabular}} &
  \begin{tabular}[c]{@{}c@{}}\texttt{r/conspiracy}\\ {\footnotesize (Conspiracy)}\end{tabular} \\ 
 &
  \multicolumn{1}{c}{\begin{tabular}[c]{@{}c@{}}\texttt{r/GenZedong}\\ {\footnotesize (Radical Left)}\end{tabular}} &
  \begin{tabular}[c]{@{}c@{}}\texttt{r/ConspiracyMemes}\\ {\footnotesize (Conspiracy)}\end{tabular} \\ 
 &
  \multicolumn{1}{c}{\begin{tabular}[c]{@{}c@{}}\texttt{r/LateStageCapitalism}\\ {\footnotesize (Radical Left)}\end{tabular}} &
  \begin{tabular}[c]{@{}c@{}}\texttt{r/environment}\\ {\footnotesize (Environment)}\end{tabular} \\ \bottomrule
\end{tabular}
\end{adjustbox}
\end{table}
\begin{table}
\centering
\caption{\textbf{Top bridge subreddits to environmental content across sociodemographic axes.}
Top three bridge subreddits ranked by personalized PageRank (PPR) for the Environment community across each sociodemographic dimension and axis. We report each subreddit's community membership below their name.}
\label{tab:environment_bridges}
\small
\begin{adjustbox}{max width=\textwidth}
\begin{tabular}{ccc}
\toprule
\multirow{2}{*}{\textbf{Social Dimension}} &
  \multicolumn{2}{c}{\textbf{Top-3 Bridges}} \\ \cmidrule(lr){2-3}
 &
  \multicolumn{1}{c}{\textbf{Negative}} &
  \textbf{Positive} \\ \midrule
\multirow{3}{*}{\textbf{\begin{tabular}[c]{@{}c@{}}Age\\ (Young - Old)\end{tabular}}} &
  \multicolumn{1}{c}{\begin{tabular}[c]{@{}c@{}}\texttt{r/Meditation}\\ {\footnotesize (Esoterism)}\end{tabular}} &
  \begin{tabular}[c]{@{}c@{}}\texttt{r/LateStageCapitalism}\\ {\footnotesize (Radical Left)}\end{tabular} \\ 
 &
  \multicolumn{1}{c}{\begin{tabular}[c]{@{}c@{}}\texttt{r/Socialism\_101}\\ {\footnotesize (Radical Left)}\end{tabular}} &
  \begin{tabular}[c]{@{}c@{}}\texttt{r/socialism}\\ {\footnotesize (Radical Left)}\end{tabular} \\ 
 &
  \multicolumn{1}{c}{\begin{tabular}[c]{@{}c@{}}\texttt{r/climateskeptics}\\ {\footnotesize (Alt-Right)}\end{tabular}} &
  \begin{tabular}[c]{@{}c@{}}\texttt{r/lostgeneration}\\ {\footnotesize (Radical Left)}\end{tabular} \\ \midrule
\multirow{3}{*}{\textbf{\begin{tabular}[c]{@{}c@{}}Gender\\ (Masculine - Feminine)\end{tabular}}} &
  \multicolumn{1}{c}{\begin{tabular}[c]{@{}c@{}}\texttt{r/ChapoTrapHouse}\\ {\footnotesize (Radical Left)}\end{tabular}} &
  \begin{tabular}[c]{@{}c@{}}\texttt{r/LateStageCapitalism}\\ {\footnotesize (Radical Left)}\end{tabular} \\ 
 &
  \multicolumn{1}{c}{\begin{tabular}[c]{@{}c@{}}\texttt{r/chapotraphouse2}\\ {\footnotesize (Radical Left)}\end{tabular}} &
  \begin{tabular}[c]{@{}c@{}}\texttt{r/Socialism\_101}\\ {\footnotesize (Radical Left)}\end{tabular} \\ 
 &
  \multicolumn{1}{c}{\begin{tabular}[c]{@{}c@{}}\texttt{r/DankLeft}\\ {\footnotesize (Radical Left)}\end{tabular}} &
  \begin{tabular}[c]{@{}c@{}}\texttt{r/FreeSpeech}\\ {\footnotesize (Alt-Right)}\end{tabular} \\ \midrule
\multirow{3}{*}{\textbf{\begin{tabular}[c]{@{}c@{}}Affluence\\ (Poor - Affluent)\end{tabular}}} &
  \multicolumn{1}{c}{\begin{tabular}[c]{@{}c@{}}\texttt{r/energy\_work}\\ {\footnotesize (Esoterism)}\end{tabular}} &
  \begin{tabular}[c]{@{}c@{}}\texttt{r/LateStageCapitalism}\\ {\footnotesize (Radical Left)}\end{tabular} \\ 
 &
  \multicolumn{1}{c}{\begin{tabular}[c]{@{}c@{}}\texttt{r/ChapoTrapHouse}\\ {\footnotesize (Radical Left)}\end{tabular}} &
  \begin{tabular}[c]{@{}c@{}}\texttt{r/Meditation}\\ {\footnotesize (Esoterism)}\end{tabular} \\ 
 &
  \multicolumn{1}{c}{\begin{tabular}[c]{@{}c@{}}\texttt{r/Anarchism}\\ {\footnotesize (Radical Left)}\end{tabular}} &
  \begin{tabular}[c]{@{}c@{}}\texttt{r/Conservative}\\ {\footnotesize (Alt-Right)}\end{tabular} \\ \midrule
\multirow{3}{*}{\textbf{\begin{tabular}[c]{@{}c@{}}Partisan\\ (Left-Wing - Right-Wing)\end{tabular}}} &
  \multicolumn{1}{c}{\begin{tabular}[c]{@{}c@{}}\texttt{r/LateStageCapitalism}\\ {\footnotesize (Radical Left)}\end{tabular}} &
  \begin{tabular}[c]{@{}c@{}}\texttt{r/Conservative}\\ {\footnotesize (Alt-Right)}\end{tabular} \\ 
 &
  \multicolumn{1}{c}{\begin{tabular}[c]{@{}c@{}}\texttt{r/ChapoTrapHouse}\\ {\footnotesize (Radical Left)}\end{tabular}} &
  \begin{tabular}[c]{@{}c@{}}\texttt{r/conspiracy}\\ {\footnotesize (Conspiracy)}\end{tabular} \\ 
 &
  \multicolumn{1}{c}{\begin{tabular}[c]{@{}c@{}}\texttt{r/socialism}\\ {\footnotesize (Radical Left)}\end{tabular}} &
  \begin{tabular}[c]{@{}c@{}}\texttt{r/canadaleft}\\ {\footnotesize (Radical Left)}\end{tabular} \\ \bottomrule
\end{tabular}
\end{adjustbox}
\end{table}
\begin{table}
\centering
\caption{\textbf{Top bridge subreddits to radical left content across sociodemographic axes.}
Top three bridge subreddits ranked by personalized PageRank (PPR) for the Radical Left community across each sociodemographic dimension and axis. We report each subreddit's community membership below their name.}
\label{tab:radical_left_bridges}
\small
\begin{adjustbox}{max width=\textwidth}
\begin{tabular}{ccc}
\toprule
\multirow{2}{*}{\textbf{Social Dimension}} &
  \multicolumn{2}{c}{\textbf{Top-3 Bridges}} \\ \cmidrule(lr){2-3}
 &
  \multicolumn{1}{c}{\textbf{Negative}} &
  \textbf{Positive} \\ \midrule
\multirow{3}{*}{\textbf{\begin{tabular}[c]{@{}c@{}}Age\\ (Young - Old)\end{tabular}}} &
  \multicolumn{1}{c}{\begin{tabular}[c]{@{}c@{}}\texttt{r/FreeSpeech}\\ {\footnotesize (Alt-Right)}\end{tabular}} &
  \begin{tabular}[c]{@{}c@{}}\texttt{r/conspiracy}\\ {\footnotesize (Conspiracy)}\end{tabular} \\ 
 &
  \multicolumn{1}{c}{\begin{tabular}[c]{@{}c@{}}\texttt{r/AstralProjection}\\ {\footnotesize (Esoterism)}\end{tabular}} &
  \begin{tabular}[c]{@{}c@{}}\texttt{r/WayOfTheBern}\\ {\footnotesize (Alt-Right)}\end{tabular} \\ 
 &
  \multicolumn{1}{c}{\begin{tabular}[c]{@{}c@{}}\texttt{r/LucidDreaming}\\ {\footnotesize (Esoterism)}\end{tabular}} &
  \begin{tabular}[c]{@{}c@{}}\texttt{r/Meditation}\\ {\footnotesize (Esoterism)}\end{tabular} \\ \midrule
\multirow{3}{*}{\textbf{\begin{tabular}[c]{@{}c@{}}Gender\\ (Masculine - Feminine)\end{tabular}}} &
  \multicolumn{1}{c}{\begin{tabular}[c]{@{}c@{}}\texttt{r/Meditation}\\ {\footnotesize (Esoterism)}\end{tabular}} &
  \begin{tabular}[c]{@{}c@{}}\texttt{r/conspiracy}\\ {\footnotesize (Conspiracy)}\end{tabular} \\ 
 &
  \multicolumn{1}{c}{\begin{tabular}[c]{@{}c@{}}\texttt{r/JordanPeterson}\\ {\footnotesize (Alt-Right)}\end{tabular}} &
  \begin{tabular}[c]{@{}c@{}}\texttt{r/WayOfTheBern}\\ {\footnotesize (Alt-Right)}\end{tabular} \\ 
 &
  \multicolumn{1}{c}{\begin{tabular}[c]{@{}c@{}}\texttt{r/Conservative}\\ {\footnotesize (Alt-Right)}\end{tabular}} &
  \begin{tabular}[c]{@{}c@{}}\texttt{r/FreeSpeech}\\ {\footnotesize (Alt-Right)}\end{tabular} \\ \midrule
\multirow{3}{*}{\textbf{\begin{tabular}[c]{@{}c@{}}Affluence\\ (Poor - Affluent)\end{tabular}}} &
  \multicolumn{1}{c}{\begin{tabular}[c]{@{}c@{}}\texttt{r/FreeSpeech}\\ {\footnotesize (Alt-Right)}\end{tabular}} &
  \begin{tabular}[c]{@{}c@{}}\texttt{r/Economics}\\ {\footnotesize (Environment)}\end{tabular} \\ 
 &
  \multicolumn{1}{c}{\begin{tabular}[c]{@{}c@{}}\texttt{r/conspiracy}\\ {\footnotesize (Conspiracy)}\end{tabular}} &
  \begin{tabular}[c]{@{}c@{}}\texttt{r/energy}\\ {\footnotesize (Environment)}\end{tabular} \\ 
 &
  \multicolumn{1}{c}{\begin{tabular}[c]{@{}c@{}}\texttt{r/ConspiracyMemes}\\ {\footnotesize (Conspiracy)}\end{tabular}} &
  \begin{tabular}[c]{@{}c@{}}\texttt{r/environment}\\ {\footnotesize (Environment)}\end{tabular} \\ \midrule
\multirow{3}{*}{\textbf{\begin{tabular}[c]{@{}c@{}}Partisan\\ (Left-Wing - Right-Wing)\end{tabular}}} &
  \multicolumn{1}{c}{\begin{tabular}[c]{@{}c@{}}\texttt{r/WayOfTheBern}\\ {\footnotesize (Alt-Right)}\end{tabular}} &
  \begin{tabular}[c]{@{}c@{}}\texttt{r/conspiracy}\\ {\footnotesize (Conspiracy)}\end{tabular} \\ 
 &
  \multicolumn{1}{c}{\begin{tabular}[c]{@{}c@{}}\texttt{r/environment}\\ {\footnotesize (Environment)}\end{tabular}} &
  \begin{tabular}[c]{@{}c@{}}\texttt{r/Meditation}\\ {\footnotesize (Esoterism)}\end{tabular} \\ 
 &
  \multicolumn{1}{c}{\begin{tabular}[c]{@{}c@{}}\texttt{r/WitchesVsPatriarchy}\\ {\footnotesize (Esoterism)}\end{tabular}} &
  \begin{tabular}[c]{@{}c@{}}\texttt{r/FreeSpeech}\\ {\footnotesize (Alt-Right)}\end{tabular} \\ \bottomrule
\end{tabular}
\end{adjustbox}
\end{table}
\begin{table}
\centering
\caption{\textbf{Top bridge subreddits to esoteric content across sociodemographic axes.}
Top three bridge subreddits ranked by personalized PageRank (PPR) for the Esoterism community across each sociodemographic dimension and axis. We report each subreddit's community membership below their name.}
\label{tab:esoterism_bridges}
\small
\begin{adjustbox}{max width=\textwidth}
\begin{tabular}{ccc}
\toprule
\multirow{2}{*}{\textbf{Social Dimension}} &
  \multicolumn{2}{c}{\textbf{Top-3 Bridges}} \\ \cmidrule(lr){2-3}
 &
  \multicolumn{1}{c}{\textbf{Negative}} &
  \textbf{Positive} \\ \midrule
\multirow{3}{*}{\textbf{\begin{tabular}[c]{@{}c@{}}Age\\ (Young - Old)\end{tabular}}} &
  \multicolumn{1}{c}{\begin{tabular}[c]{@{}c@{}}\texttt{r/alltheleft}\\ {\footnotesize (Radical Left)}\end{tabular}} &
  \begin{tabular}[c]{@{}c@{}}\texttt{r/conspiracy}\\ {\footnotesize (Conspiracy)}\end{tabular} \\ 
 &
  \multicolumn{1}{c}{\begin{tabular}[c]{@{}c@{}}\texttt{r/EscapingPrisonPlanet}\\ {\footnotesize (Conspiracy)}\end{tabular}} &
  \begin{tabular}[c]{@{}c@{}}\texttt{r/AntifascistsofReddit}\\ {\footnotesize (Radical Left)}\end{tabular} \\ 
 &
  \multicolumn{1}{c}{\begin{tabular}[c]{@{}c@{}}\texttt{r/conspiracytheories}\\ {\footnotesize (Conspiracy)}\end{tabular}} &
  \begin{tabular}[c]{@{}c@{}}\texttt{r/ConspiracyMemes}\\ {\footnotesize (Conspiracy)}\end{tabular} \\ \midrule
\multirow{3}{*}{\textbf{\begin{tabular}[c]{@{}c@{}}Gender\\ (Masculine - Feminine)\end{tabular}}} &
  \multicolumn{1}{c}{\begin{tabular}[c]{@{}c@{}}\texttt{r/LateStageCapitalism}\\ {\footnotesize (Radical Left)}\end{tabular}} &
  \begin{tabular}[c]{@{}c@{}}\texttt{r/conspiracy}\\ {\footnotesize (Conspiracy)}\end{tabular} \\ 
 &
  \multicolumn{1}{c}{\begin{tabular}[c]{@{}c@{}}\texttt{r/SocialistRA}\\ {\footnotesize (Radical Left)}\end{tabular}} &
  \begin{tabular}[c]{@{}c@{}}\texttt{r/EscapingPrisonPlanet}\\ {\footnotesize (Conspiracy)}\end{tabular} \\ 
 &
  \multicolumn{1}{c}{\begin{tabular}[c]{@{}c@{}}\texttt{r/ChapoTrapHouse}\\ {\footnotesize (Radical Left)}\end{tabular}} &
  \begin{tabular}[c]{@{}c@{}}\texttt{r/ConspiracyMemes}\\ {\footnotesize (Conspiracy)}\end{tabular} \\ \midrule
\multirow{3}{*}{\textbf{\begin{tabular}[c]{@{}c@{}}Affluence\\ (Poor - Affluent)\end{tabular}}} &
  \multicolumn{1}{c}{\begin{tabular}[c]{@{}c@{}}\texttt{r/conspiracy}\\ {\footnotesize (Conspiracy)}\end{tabular}} &
  \begin{tabular}[c]{@{}c@{}}\texttt{r/AntifascistsofReddit}\\ {\footnotesize (Radical Left)}\end{tabular} \\ 
 &
  \multicolumn{1}{c}{\begin{tabular}[c]{@{}c@{}}\texttt{r/ConspiracyMemes}\\ {\footnotesize (Conspiracy)}\end{tabular}} &
  \begin{tabular}[c]{@{}c@{}}\texttt{r/JordanPeterson}\\ {\footnotesize (Alt-Right)}\end{tabular} \\ 
 &
  \multicolumn{1}{c}{\begin{tabular}[c]{@{}c@{}}\texttt{r/alltheleft}\\ {\footnotesize (Radical Left)}\end{tabular}} &
  \begin{tabular}[c]{@{}c@{}}\texttt{r/ZeroWaste}\\ {\footnotesize (Environment)}\end{tabular} \\ \midrule
\multirow{3}{*}{\textbf{\begin{tabular}[c]{@{}c@{}}Partisan\\ (Left-Wing - Right-Wing)\end{tabular}}} &
  \multicolumn{1}{c}{\begin{tabular}[c]{@{}c@{}}\texttt{r/lostgeneration}\\ {\footnotesize (Radical Left)}\end{tabular}} &
  \begin{tabular}[c]{@{}c@{}}\texttt{r/conspiracy}\\ {\footnotesize (Conspiracy)}\end{tabular} \\ 
 &
  \multicolumn{1}{c}{\begin{tabular}[c]{@{}c@{}}\texttt{r/LateStageCapitalism}\\ {\footnotesize (Radical Left)}\end{tabular}} &
  \begin{tabular}[c]{@{}c@{}}\texttt{r/JordanPeterson}\\ {\footnotesize (Alt-Right)}\end{tabular} \\ 
 &
  \multicolumn{1}{c}{\begin{tabular}[c]{@{}c@{}}\texttt{r/Anarchism}\\ {\footnotesize (Radical Left)}\end{tabular}} &
  \begin{tabular}[c]{@{}c@{}}\texttt{r/ConspiracyMemes}\\ {\footnotesize (Conspiracy)}\end{tabular} \\ \bottomrule
\end{tabular}
\end{adjustbox}
\end{table}

\endgroup

\end{document}